\documentclass[twocolumn]{aastex62}
\usepackage{graphicx}   
\usepackage{subfigure}
\usepackage{multirow}
\usepackage{amsmath}

\graphicspath{{./}{figures/}}

\received{XXX}
\revised{XXX}
\accepted{XXX}
\submitjournal{AAS journal}

\shorttitle{The dispersion measure and scattering of FRBs}
\shortauthors{Zhu \& Feng}

\begin{document}

\title{The dispersion measure and scattering of FRBs: Contributions from the intergalactic medium, foreground halos, and hosts}

\author{Weishan, Zhu}
\affil{School of Physics and Astronomy, Sun Yat-Sen University, Zhuhai campus, No. 2, Daxue Road \\
Zhuhai, Guangdong, 519082, China}

\author{Long-Long, Feng}
\affil{School of Physics and Astronomy, Sun Yat-Sen University, Zhuhai campus, No. 2, Daxue Road \\
Zhuhai, Guangdong, 519082, China}
\affiliation{Purple Mountain Observatory, CAS, No.10 Yuanhua Road, Qixia District, Nanjing 210033, China}



\begin{abstract}
We investigate the dispersion measure(DM) and scattering of FRBs by the intergalactic-medium(IGM), foreground and host halos, using cosmological hydrodynamical simulation. We find that the median DM caused by foreground halos is around 30\% of that caused by the IGM, but has a much larger variance. The DM induced by hosts deviates from a log-normal distribution, but exhibits an extended distribution in the range of $1-3000 \ {\rm{pc\, cm^{-3}}}$ with a median value $\sim 100 \ {\rm{pc\, cm^{-3}}}$. Then we produce mock FRB sources, assuming a uniform distribution in the range $z\sim 0-0.82$, to consider the propagation effect of IGM, foreground and host halos on FRB signals simultaneously. The DM distribution of mock sources agrees well with the observation. The fitted DM-redshift relation of the mock sources can provide a rough estimation of the redshifts of observed events with errors $\delta z \lesssim 0.15$. The distribution of mock sources in the DM-scattering time($\tau$) space can also match the observation, assuming a Kolmogorov turbulence model with the inner and outer scale is 1000 km to 1 AU, and 0.2-10 pc respectively. Finally, we estimate the relative importance of these medium on DM and $\tau$ in our models. The IGM and host halos are the primary and secondary sources to the extragalactic DM, $\rm{DM_{exg}}$. Meanwhile, the contribution from foreground halos increases as $\rm{DM_{exg}}$ increases. The host and foreground halos may be the most important medium for scattering. Statistically, the latter may dominate the scattering of events with $\rm{DM_{exg}} \gtrsim 200 \ {\rm{pc\, cm^{-3}}}$. 
\end{abstract}

\keywords{radio continuum: general --- intergalactic medium---galaxies:halos--- turbulence  }


\section{Introduction} \label{sec:intro}
Fast radio bursts (FRBs) are a class of bright and millisecond-duration radio transients at cosmological distances. The first event was reported in 2007(\citealt{2007Sci...318..777L}). In the past few years, the number of detected events has been increasing sharply, mainly thanks to the Parkes telescope, UTMOST telescopes, ASKAP telescope and the CHIME cylinder array. By the middle 2020, more than 100 sources have been reported (e.g., \citealt{2013Sci...341...53T};  \citealt{2016PASA...33...45P};  \citealt{2018MNRAS.475.1427B};  \citealt{2018Natur.562..386S};  \citealt{2019Natur.566..230C}), and 19 of them are repeating sources(\citealt{2016Natur.531..202S};  \citealt{2019Natur.566..235C};  \citealt{2019ApJ...885L..24C}; \citealt{2020ApJ...891L...6F}). Recently, periodic activity has been reported for the repeating event FRB  180916.J0158+65 (\citealt{2020arXiv200110275T}). So far, eight events have been localized to their host galaxies(\citealt{2017ApJ...834L...7T}; \citealt{2019Sci...365..565B}; \citealt{2019Natur.572..352R}; \citealt{2020Natur.577..190M}; \citealt{2020Natur.581..391M}). Based on observations, the all sky events rate is expected to be a few to tens of thousands per day with a fluence above $\rm{1\ Jy\ ms}$(e.g.,  \citealt{2013Sci...341...53T}; \citealt{2014ApJ...790..101S}; \citealt{2015MNRAS.447.2852K}). The physical origin of FRBs is still unknown, although many models have been proposed(see \citealt{2019PhR...821....1P} for a review). Very recently, the Galactic magnetar SGR 1935+2154 emitted a millisecond-duration radio burst with a spectral energy that is about 40 times less than the weakest FRB, suggesting that active magnetar might be able to produce FRBs at cosmological distances(\citealt{2020arXiv200510828B}).


During the travel from sources to observers, FRBs signals would pass through all kinds of intervening material, including medium surrounding the sources, the interstellar medium(ISM) and the circumgalactic medium (CGM) of host galaxies, the diffuse intergalactic medium (IGM), the gaseous halos of foreground galaxies, and the interstellar medium of the Milky Way. These materials would cause many propagation effects and lead to several important features in the observed pulse signal(\citealt{2019A&ARv..27....4P}; \citealt{2019ARA&A..57..417C}). Dispersion and scattering of FRBs, as well as polarization, have been measured for all or some of the events. The dispersion measures (DMs) of reported FRBs range from slightly over 100 to 2600 ${\rm{pc\, cm^{-3}}}$, and are larger than the expected value contributed by the Milk Way, $\rm{DM_{MW}}$, and its halo. $\rm{DM_{MW}}$ is usually estimated by the models in \cite{2002astro.ph..7156C} and \cite{2017ApJ...835...29Y}, known as NE2001 and YMW16 respectively, while the contribution from Galactic halo is taken to be $\approx 30-50\ \rm{pc\, cm^{-3}}$(e.g. \citealt{2016arXiv160505890C}). Because of the excess of DM, FRBs are identified as extragalactic sources. 

For most of the FRBs, their distance/redshift are estimated by the dispersion measure induced by extragalactic gas, denoted as $\rm{DM_{exg}}$. This procedure usually assumes different models of the IGM distribution and the DMs caused by hosts, i.e., $\rm{DM_{host}}$, to obtain the contribution from the IGM, denoted as $\rm{DM_{IGM}}$(e.g., \citealt{2014ApJ...783L..35D}). Then, the distance/redshift can be inferred from the $\rm{DM_{IGM}}$-redshift relation that have been constructed by analytical and simulation works(e.g., \citealt{2003ApJ...598L..79I};\citealt{2004MNRAS.348..999I}; \citealt{2014ApJ...780L..33M}; \citealt{2015MNRAS.451.4277D}). The derived distance/redshift can help to localize the host galaxies(\citealt{2017ApJ...849..162E}), and estimate the luminosity function of FRBs(e.g., \citealt{2018MNRAS.481.2320L}). On the other hand, observations of a large amount of FRBs from different redshifts could be used in turn to  probe the distribution of baryons in the universe(e.g., \citealt{2016ApJ...830...75V};  \citealt{2018ApJ...852L..11S}; \citealt{2020Natur.581..391M}), and constrain the cosmology parameters(e.g., \citealt{2016ApJ...830L..31Y}; \citealt{2018ApJ...856...65W}). In addition, combined with the scattering and rotation measure, DMs of FRBs can further probe the magnetic field and clumpiness in the IGM(e.g \citealt{2016ApJ...824..105A}; \citealt{2018MNRAS.480.3907V}), and constrain the physical origins of FRBs(see \citealt{2019PhR...821....1P} and reference therein).

Yet, there are significant uncertainties in current models and assumptions about the contribution to the dispersion measure and scattering of FRBs from different components(\citealt{2019ARA&A..57..417C}). As for the DM, there are barely any solid observational constraint on the contribution from host galaxies and their CGM so far. Theoretical models suggest that the DM caused by a Milky Way like host disk galaxy may follow a log-normal distribution, with a median value of $\sim 100 \rm{cm^{-3} pc}$. However, different inclination angle and galaxy morphology types can lead to striking variations in $\rm{DM_{host}}$ (e.g., \citealt{2015RAA....15.1629X}; \citealt{2018arXiv180401548W}; \citealt{2018MNRAS.481.2320L}). In addition, the inhomogeneity in the IGM, especially gas in filaments and gaseous halos, could lead to considerable variations on $\rm{DM_{IGM}}$ along different line-of-sights(e.g., \citealt{2014ApJ...780L..33M}; \citealt{2018ApJ...865..147Z}; \citealt{2019ApJ...886..135P}). Moreover, the line-of-sights toward some FRB events could have passed through the gaseous halos of foreground galaxies(\citealt{2016arXiv160505890C}; \citealt{2019Sci...366..231P}; \citealt{2020arXiv200201399C}), which, however, is not well studied.

For the scattering measure, the primary contribution is also from extragalactic medium. However, the relative importance of host halos, the IGM and foreground halos is also unclear. If $\rm{DM_{host}}$ constitutes a considerable fraction of $\rm{DM_{exg}}$, for instance comparable to or more than $\sim 20\%$, the host galaxies may cause enough time broadening to explain the observed FRBs events(\citealt{2016arXiv160505890C}, \citealt{2016ApJ...832..199X}). The scattering caused by the diffuse IGM with density close to the cosmic mean is likely very weak(\citealt{2013ApJ...776..125M}, \citealt{2016ApJ...818...19K};\citealt{2014ApJ...785L..26L}; \citealt{2018ApJ...865..147Z}). However, the electrons in the highly overdense region of the IGM, such as cosmic filaments and gaseous halos, may induce significant scattering(\citealt{2013ApJ...776..125M}, \citealt{2018ApJ...865..147Z}). Due to the uncertainties in the role of host galaxies, foreground halos and the IGM, currently, the power to probe the IGM, and constrain physical origin of FRBs and cosmology using DM and scattering of FRBs is largely limited.

In this work, we make use of cosmological hydrodynamical simulations  with adaptive mesh refinement to study the contributions to the dispersion measure and scattering of FRBs by the IGM, gaseous halos of foreground galaxies, and host halos. The organization of this paper is as follows: Section 2 gives a brief description of the simulation and the numerical methods used in this work. Section 3 investigates the dispersion measure and scattering measure caused by the IGM, foreground halos, and host halos, and their statistical distributions. In section 4, we introduce the mock FRB sample and the scattering models, and further investigate the DM distribution, DM-redshift and DM-$\tau$ relations. We also make a quantitative study of the relative importance of different types of intervening medium on the total DM and time broadening of FRBs in our models. Finally, we summary our findings and discuss the application and limitation of this work in Section 5.  

\section{Methodology}

\subsection{Simulation}
We use the cosmological hydrodynamical simulation code RAMSES(\citealt{2002A&A...385..337T}) to track the evolution of cosmic matter in a cubic box with side length of $100 h^{-1}$ Mpc. We adopt a $\Lambda$CDM cosmology with parameters  $\Omega_{m}=0.317, \Omega_{\Lambda}=0.683,h=0.671,\sigma_{8}=0.834, \Omega_{b}=0.049$, and $n_{s}=0.962$(\citealt{2014A&A...571A..16P}). The simulation contains $1024^3$ dark matter particles, corresponding to a mass resolution of $1.03 \times \rm{10^{8}M_{\odot}}$. We use a number of $1024^3$ root grid and the maximum level of adaptive mesh refinement is $l_{max}=17$. The spacial resolution of the coarse and finest grid are $97.6 h^{-1}$kpc and $0.763 h^{-1}$kpc respectively. When the number of dark matter particles in a grid cell is greater than 8, this grid cell will be refined. This simulation starts at $z=99$ and ends at $z=0$. At redshift $z=8.5$, a uniform UV background according to the model in \cite{1996ApJ...461...20H} is switched on. Gas cooling, star formation and stellar feedback are implemented, but super massive black hole and AGN feedback are not modeled. Star formation is triggered in regions where the number density of hydrogen exceeds $0.1\, \rm{cm^{-3}}$. 10\% of the newly formed stars would explode as supernovae, by assuming a Salpeter initial mass function. An amount of $10^{51}$ erg energy is injected to the gas by each supernova, of which the average progenitor mass is assumed to be 10 $M_{\odot}$. The stellar feedback is implemented using the feedback module in RAMSES. 

A total number of 44 snapshots are stored during the simulation. We search for dark matter halos in snapshots after simulation using the HOP algorithm(\citealt{1998ApJ...498..137E}).  For the sake of reliability, we only consider halos that have more than 1200 dark matter particles, i.e, $\rm{M_h>1.2 \times 10^{11}M_{\odot}}$, in the following study. We identify about 35000-40000 dark matter halos with $\rm{M_h>1.2 \times 10^{11}M_{\odot}}$ in our snapshots at low redshifts.

\begin{figure}[htbp]
\begin{center}
\includegraphics[width=1.20\columnwidth, trim= 110 40 40 40, clip]{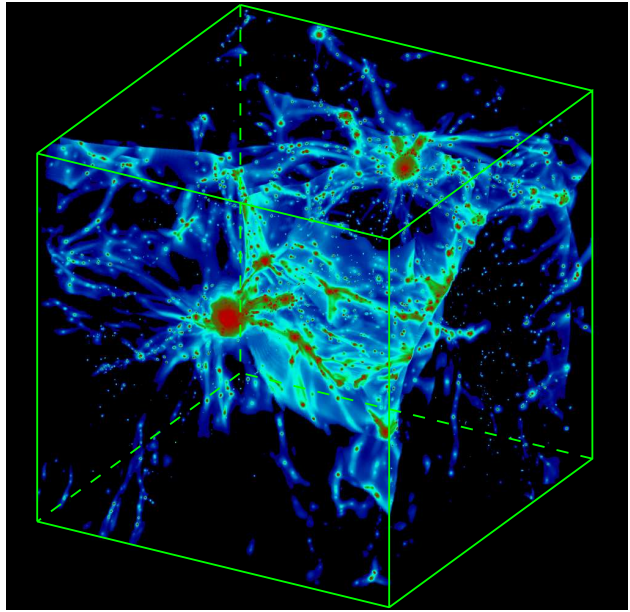}
\caption{The distribution of baryonic gas in a cubic box with size of $(25 h^{-1} \rm{Mpc})^3$ at $z=0$. }
\end{center}
\label{fig:gas_den_exam}
\end{figure}

\subsection{Calculation of DM and scattering measure}
Fig.~\ref{fig:gas_den_exam} shows the projected gas density around a massive halo at $z=0.03$ in our simulation. It clearly indicates that there are many gaseous clumps with various size residing in the diffuse IGM. These clumps are associated with halos with different masses. It is expected that some of the light paths toward FRB events would pass through some of these gaseous clumps. \cite{2016arXiv160505890C} has included such kind of clumps into their models of the dispersion and scattering of FRB events. Here, we decompose the dispersion and scattering of FRBs contributed by the intervening medium along the lines-of-sight(LOS) into three parts as the medium in the host halos, the gaseous halos of foreground galaxies that had been passed through, and the IGM but excluding the medium in the second term. We will refer to them shortly as the host halos, the foreground halos and the IGM hereafter. We evaluate the DM and scattering measure brought by these three components according to procedures described in following subsections. Note that, since we cannot resolve the gas density below $\sim 1$ kpc, we will not take the effect of the local medium surrounding the sources into account. 

\subsubsection{The IGM and foreground halos}

For a FRB source residing at redshift $z_f$, the sum of the DM contributed by the IGM and foreground halos, denoted as $\rm{DM_{IGM}}$ and $\rm{DM_{halos}}$ respectively, is given by(\citealt{2014ApJ...780L..33M}; \citealt{2014ApJ...783L..35D})
\begin{equation}
\label{eq:dm_igm_halos}
\begin{aligned}
{\rm{DM_{IGM}}}(z_f)+{\rm{DM_{halos}}}(z_f)=\int_{0}^{z_f}\frac{n_e(z)}{1+z}dl,
\end{aligned}
\end{equation}
 where $n_e(z)$ is the number density of electrons at $z$ along the LOS. In this work, we assume that all the gaseous intervening medium are fully ionized, and the density power spectrum of these inhomogeneous medium along the LOS toward FRBs follows the form
\begin{equation}
    P(k)=C_N^2k^{-\beta}e^{-kl_0}, k>L_0^{-1},
\end{equation}
in the turbulent range, where $L_0$ and $l_0$ are the outer and inner scale of turbulence. For medium with a index $\beta>3$, $C_N^2(z)$ can be approximately related to the density variance of electron$\langle\delta n_e^2(z)\rangle$ as
\begin{equation}
C_N^2(z) \approx \frac{\beta-3}{2(2\pi)^{4-\beta}} \langle \delta n_e^2(z) \rangle L_0^{3-\beta}.
\end{equation}
Consequently, the sum of the effective scattering measure due to the IGM and foreground halos would be(e.g., \citealt{2013ApJ...776..125M}; \citealt{2016ApJ...832..199X})
\begin{equation}
\label{eq:smeff_short}
\begin{aligned}
{\rm{SM}}_{\rm{eff,IGM}}(z_f)+{\rm{SM}}_{\rm{eff,halos}}(z_f)=\int_{0}^{z_f}\frac{C_N^2(z)d_H(z)}{(1+z)^3}dz, 
\end{aligned}
\end{equation}
where $d_H(z)$ indicates the Hubble radius. 
We further assume that the turbulence in the IGM, foreground and host halos fulfill the Kolmogorov turbulence model, i.e., $\beta=11/3$, and take $ \langle \delta n_e^2(z) \rangle \sim n_e^2(z)$. Therefore, eqn.~\ref{eq:smeff_short} can be expanded as 
\begin{equation}
\label{eq:smeff_long}
\begin{aligned}
{\rm{SM}}_{\rm{eff,IGM}}(z_f)+{\rm{SM}}_{\rm{eff,halos}}(z_f) &\\
\approx 1.42\times10^{-13} (\frac{\Omega_b}{0.049})^2(\frac{L_0}{1 \rm{pc}})^{-2/3}
\rm{m}^{-20/3}\\ \times \int_{0}^{z_f} (\rho_b(z)/\bar{\rho}_b(z))^2(1+z)^3d_H(z)dz &\\
\approx 1.31\times 10^{13} \rm{m}^{-17/3} \cdot \frac{1}{h}  (\frac{\Omega_b}{0.049})^2(\frac{L_0}{1 \rm{pc}})^{-2/3}\\
 \times \int_{0}^{z_f} (\rho_b(z)/\bar{\rho}_b(z))^2 \frac{(1+z)^3}{[\Omega_{\Lambda}+\Omega_m(1+z)^3]^{1/2}}dz.
\end{aligned}
\end{equation}

We adopt following procedures to evaluate $\rm{DM_{IGM}}$, $\rm{DM_{halos}}$, $\rm{SM_{eff,IGM}}$, and $\rm{SM_{eff,halos}}$.
From the outputs of simulation at different redshifts, the gas density on an uniform $4096^3$ grid, corresponding to a grid level of 12 in simulation, are constructed. Grids at this level have a spacial resolution of $24.4 h^{-1}$kpc. For each grid cell at different snapshots, we check whether it is within any dark matter halo more massive than $1.2 \times 10^{11}M_{\odot}$ or not. If a cell locate in a halo, we flag it as a cell of foreground halos. Otherwise, this cell will be flagged as a cell of the IGM.

Following conventional ways (e.g., \citealt{2001ApJ...551....3G}; \citealt{2015MNRAS.451.4277D}), we stack the gas density on the $4096^3$ grids of simulation box at different redshifts with rotating and flipping, and construct lines-of-sight to perform the integration in Eqn.~\ref{eq:smeff_long}. 
For $\rm{DM_{halos}}$ and $\rm{SM_{eff, halos}}$, the integration is only performed for those cells belonging to foreground halos. Meanwhile, only those cells belonging to the IGM are taken into accounted when measuring $\rm{DM_{IGM}}$ and $\rm{SM_{eff, IGM}}$. Due to the limited number of snapshots outputed from our cosmological simulation, the stacking and integration is ended at the redshift $z=0.82$. At higher redshifts, we lack enough snapshots to sample the distribution of the IGM and foreground halos continuously.

\subsubsection{Host halos}
If there is a FRB source siting at the center of a halo at $z_h$,  the DM caused by the host halo is evaluated as,
\begin{equation}
\label{eq:dm_hosts}
\begin{aligned}
{\rm{DM_{host}}}=\int_{0}^{r200}\frac{n_e(r)}{1+z_h}dr=\frac{1}{1+z_h}\int_{0}^{r200}n_e(r)dr,
\end{aligned}
\end{equation}
where $r200$ is the virial radius of the dark matter halo. The scattering caused by the halo gas is defined as,
\begin{equation}
\label{eq:smeff_hosts}
\begin{aligned}
{\rm{SM}}_{\rm{eff,host}}(z_h)
\approx 1.31\times 10^{13} m^{-17/3} \cdot \frac{1}{h}  (\frac{\Omega_b}{0.049})^2(\frac{L_0}{1 \rm{pc}})^{-2/3}\\
 \times \int_{0}^{r200} (\rho_b(r)/\bar{\rho}_b(z_h))^2 \frac{(1+z_h)^3}{[\Omega_{\Lambda}+\Omega_m(1+z_h)^3]^{1/2}}dr.
\end{aligned}
\end{equation}
For each dark matter halo with mass $M_h>1.2\times 10^{11}M_{\odot}$ in different snapshots, we randomly selecting 40 radial paths from the halo center to the halo boundary to produce 40 sets of $\rm{DM_{host}}$ and $\rm{SM_{eff,host}}$. Each radial path is evenly divided into to 50 segments. The baryon density at each segment is evaluated from the distribution of baryonic gas in the halos that resolved in our AMR simulation. By virtue of the AMR technique, the density field in halos can be resolved on fine grids with spacial resolution of a few kpc. Then if the total number and the redshift distribution function of mock FRBs are given, we will assign these mock FRBs to halos in different snapshots. Each mock FRB is randomly associated to one of these 40 radial paths of the assigned halo, and corresponding $\rm{DM_{host}}$ and $\rm{SM_{host}}$. 

\begin{figure}[htbp]
\begin{center}
\includegraphics[width=0.90\columnwidth]{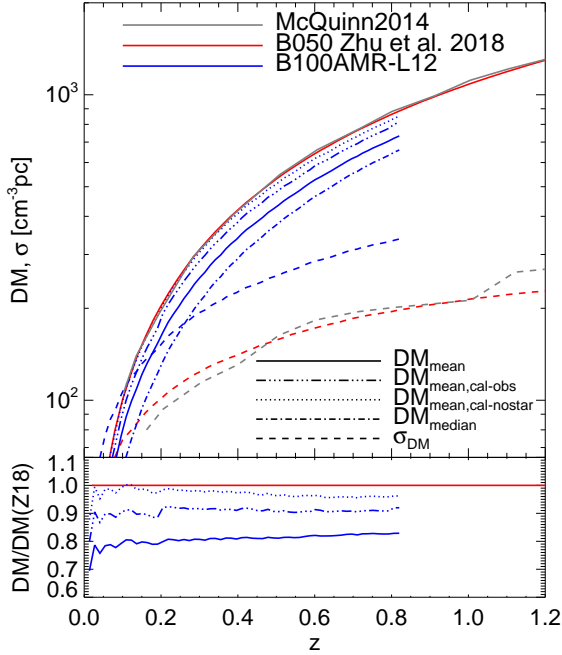}
\vspace{-0.3cm}
\caption{Top: Blue solid, dotted-dashed, and short dashed lines indicate the mean, median and standard deviation of $\rm{DM}$ caused by the IGM and foreground halos as a function of the redshift, based on the simulation in this work. The triple-dotted-dashed line indicates the expected mean $\rm{DM}$, if the stellar density is calibrated by observation. The dotted line shows the expected mean $\rm{DM}$, if there is no star formation. The lines in color of gray and red indicate the results in \citet{2014ApJ...780L..33M} and \citet{2018ApJ...865..147Z} respectively. Bottom: The ratio of the mean $\rm{DM}$ under three different cases of star density to the result in \citet{2018ApJ...865..147Z}.}
\end{center}
\label{fig:dm_z}
\end{figure}

\begin{figure}[htbp]
\begin{center}
\includegraphics[width=0.90\columnwidth]{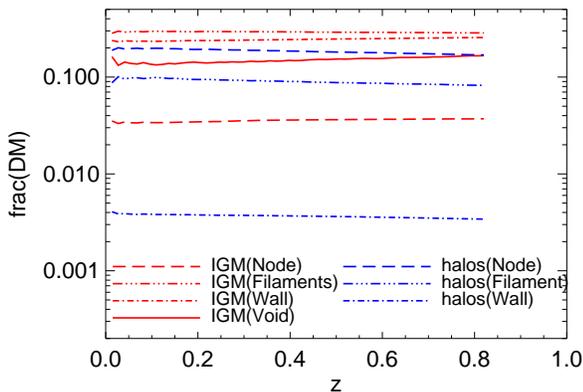}
\vspace{-0.3cm}
\caption{Red lines indicate the contributions to $\rm{DM_{IGM}(z)+DM_{halos}(z)}$ from the IGM residing in nodes(dashed), filaments(triple-dotted-dashed), walls(dotted-dashed) and voids(solid), respectively. Blue lines indicate the contributions from the foreground halos in corresponding structures.}
\end{center}
\label{fig:dm_frac}
\end{figure}

\begin{figure*}[htbp]
\begin{center}
\hspace{-0.0cm}
\includegraphics[angle=90,width=1.00\textwidth]{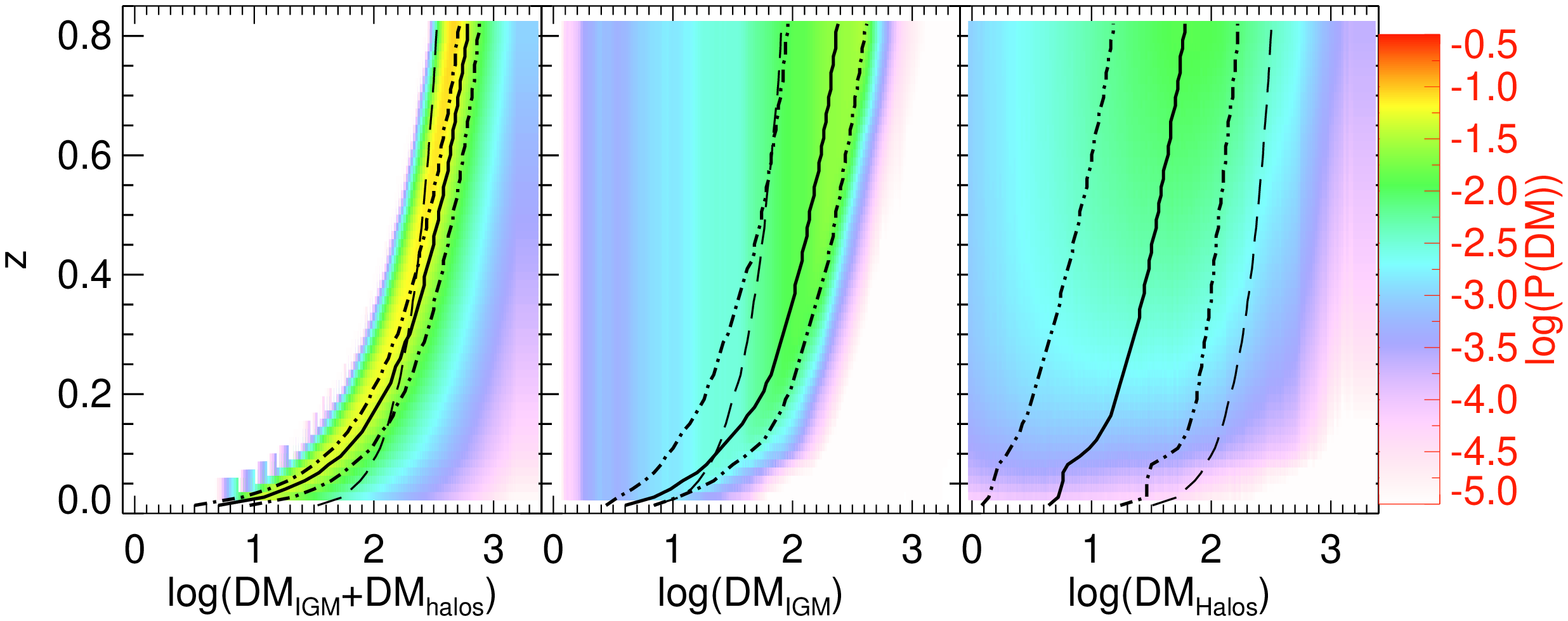}
\caption{Left, middle and right panel show the probability density distribution of $\rm{DM_{IGM}(z)+DM_{halos}(z)}$,  $\rm{DM_{IGM}(z)}$, and $\rm{DM_{halos}(z)}$ respectively. Solid line in each panel indicates the median value of DM at different z. The left and right dotted-dashed lines indicate the $20\%$ and $80\%$ percentiles of DM at different z. The dashed line indicates the standard deviation. }
\end{center}
\label{fig:dm_pdf}
\end{figure*}

\section{DM and SM caused by the IGM, foreground halos and host halos}
In this section, we introduce our results on the DM and SM induced by the IGM, foreground halos and host halos, based on the methods described in the last section.
\subsection{DM of the IGM and foreground halos}
We randomly sampled one million lines of sight to estimate the DM and SM caused by the IGM and foreground halos. Due to the limited number of snapshots produced by our simulation, the LOSs are ended at $z=0.82$. In the top panel of of Fig.~\ref{fig:dm_z}, we show the mean and median value of ${\rm{DM}}$ caused by the IGM and foreground halos as a function of the redshift, i.e., $\rm{DM_{IGM}(z)+DM_{halos}(z)}$, and the corresponding standard deviation at $0<z<0.82$ by solid, dotted-dashed and short-dashed lines, respectively. In comparison, the results in \cite{2014ApJ...780L..33M}(M14) and \cite{2018ApJ...865..147Z}(Z18) are shown by gray and red lines respectively. Note that, the contribution from foreground halos is not separated from the contribution of the IGM in M14 and Z18. 

The magnitude of $\rm{DM_{IGM}(z)+DM_{halos}(z)}$ in this work is lower than the corresponding result in Z18 by about $20\%$. The major reason is that star formation is modeled in this work but not included in Z18. However, the stellar component is somehow over produced in our AMR simulation, probably due to the lack of AGN feedback. If we lower down the mass fraction of star by hand, i.e., calibrate it by the observed cosmic star density(\citealt{2008ApJ...675..234P}), $\rm{DM_{IGM}(z)+DM_{halos}(z)}$ could be increased to $\sim 90\%$ of Z18. If we further assume that there was no star formation, the expected mean value of $\rm{DM_{IGM}(z)+DM_{halos}(z)}$ would be slightly lower than Z18, and is about $\sim 98\%-100\%$ of Z18 at $z<0.2$, and $\sim90\%-95\%$ of Z18 at $0.2<z<0.82$. This minor discrepancy might result from variance along LOSs, and the different number of LOSs and different simulation resolution. 

Inferred from our simulation, the dependence of the median value of $\rm{DM_{IGM}(z)+DM_{halos}(z)}$ is found to be well fitted by
\begin{equation}
\label{eq:dm_z_fit}
\begin{aligned}
\rm{log_{10}DM(z)}=2.92+1.13\rm{log}_{10}(z)-0.03(\rm{log}_{10}(z))^2. 
\end{aligned}
\end{equation}
While for the mean values of $\rm{DM_{IGM}(z)+DM_{halos}(z)}$, we found that it can adopt a similar form of fitting formula but with an alternative set of coefficients (2.96, 1.07, 0.01) on the right hand side of equation. In comparison, the corresponding coefficients are (3.02, 1.00, -0.01) while fitting the results in Z18 with the similar formula. 

In Fig.~\ref{fig:dm_frac}, we show the contributions to the sum of $\rm{DM_{IGM}(z)+DM_{halos}(z)}$ from matter in different structures. On average, foreground halos approximately accounts for $\sim 20-30\%$ of $\rm{DM_{IGM}(z)+DM_{halos}(z)}$, and the rest is caused by the IGM. As in Z18, we further assign the grid cells into four categories of cosmic large scale structures, using the method given in \cite{2017ApJ...838...21Z}. At $z=0$, halos residing in nodes/clusters, filaments and walls contributes $\sim 20\%$, $\sim 10\%$ and $\sim 0.4\%$ of $\rm{DM_{IGM}(z)+DM_{halos}(z)}$ respectively. Those fractions decrease gradually as redshift increases. In contrast, the IGM residing in nodes, filaments, walls and voids contributes $\sim 3.5\%$, $\sim 28\%$, $\sim 23\%$ and $\sim 15 \%$ of $\rm{DM_{IGM}(z)+DM_{halos}(z)}$ respectively at $z=0$.

The standard deviation of DM in our AMR simulation is around 1.5 times of that in Z18 and M14. This feature should be due to the higher resolution in this work. The spacial resolution at the refinement level of 12 in our AMR simulation is $24.4 h^{-1}$kpc, while the simulation B050 in Z18 has a resolution of $48.8 h^{-1}$kpc. In Z18, we have demonstrated that relatively poor resolution will underestimate the baryon density and mass fraction in highly over-dense region, leading to a lower standard deviation of DM. On the other hand, this work shows that the standard deviation of DM contributed by the IGM and foreground halos could be larger than 200 ${\rm{pc\, cm^{-3}}}$ at $z>0.2$. 

Based on the matter density in the simulation Illustris-3, \cite{2019MNRAS.484.1637J} obtained a value of $\sigma(\rm{DM_{IGM}})=115 {\rm{pc\, cm^{-3}}}$ at $z=1$, and $\sigma(\rm{DM_{IGM}+DM_{hlaos}}) \sim {\rm{200 pc\, cm^{-3}}}$. The latter is consistent with \citet{2014ApJ...780L..33M} and \citet{2018ApJ...865..147Z}, but is smaller than the result at $z=0.80$ of this work. This should be mainly caused by the difference of resolutions among these simulations. Illustris-3 provides the density of $455^3$ gas cells in a cubic box with side length $75/h$ Mpc, which may underestimate the variance among different lines of sights due to limited resolution. Based on the halo model, \cite{2019MNRAS.485..648P} demonstrated that there are large scatters in the DM caused by foreground halo gas, depending highly on the impact parameter and halo mass. Particularly, the larger scatters are attributed to massive halos, because galaxy group and clusters can give rise to DM as large as several hundreds to thousands of ${\rm{pc\, cm^{-3}}}$.

To illustrate the distribution of $\rm{DM_{IGM}}$ and $\rm{DM_{halos}}$ as a function of redshift, and their contribution to the significant variance of $\rm{DM_{IGM}(z)+DM_{halos}(z)}$ among different LOSs, we present the probability density distributions in Fig.~\ref{fig:dm_pdf}. We find that both the IGM and gaseous halos of foreground galaxies can give rise to a significant variance. But the variance caused by the gaseous halos of foreground galaxies is much larger than that by the IGM. In the redshift range we studied, the standard deviation of $\rm{DM_{halos}}$ can be larger than that of $\rm{DM_{IGM}}$ by a factor of $\sim 3-4$, despite that the median value of $\rm{DM_{halos}}$ is only about one thirds of the median value of $\rm{DM_{IGM}}$. At $z=0.82$, $\sigma(\rm{DM_{IGM}}) \simeq 80 {\rm{pc\, cm^{-3}}}$, while $\sigma(\rm{DM_{halos}}) \simeq 300 {\rm{pc\, cm^{-3}}}$. This striking feature results from different lines-of-sights having different impact path to halos of foreground galaxies, and those halos differ from each other significantly in properties such as mass, radius and density distribution. Our result is basically agreement with \cite{2019MNRAS.485..648P}. This result suggests that it is important to verify whether the line-of-sight to a FRB event have passed through any foreground halo or not, in order to precisely estimate the redshift of this event from its DM.

\begin{figure}[htbp]
\begin{center}
\includegraphics[width=0.80\columnwidth,trim=0 25 0 10,clip]{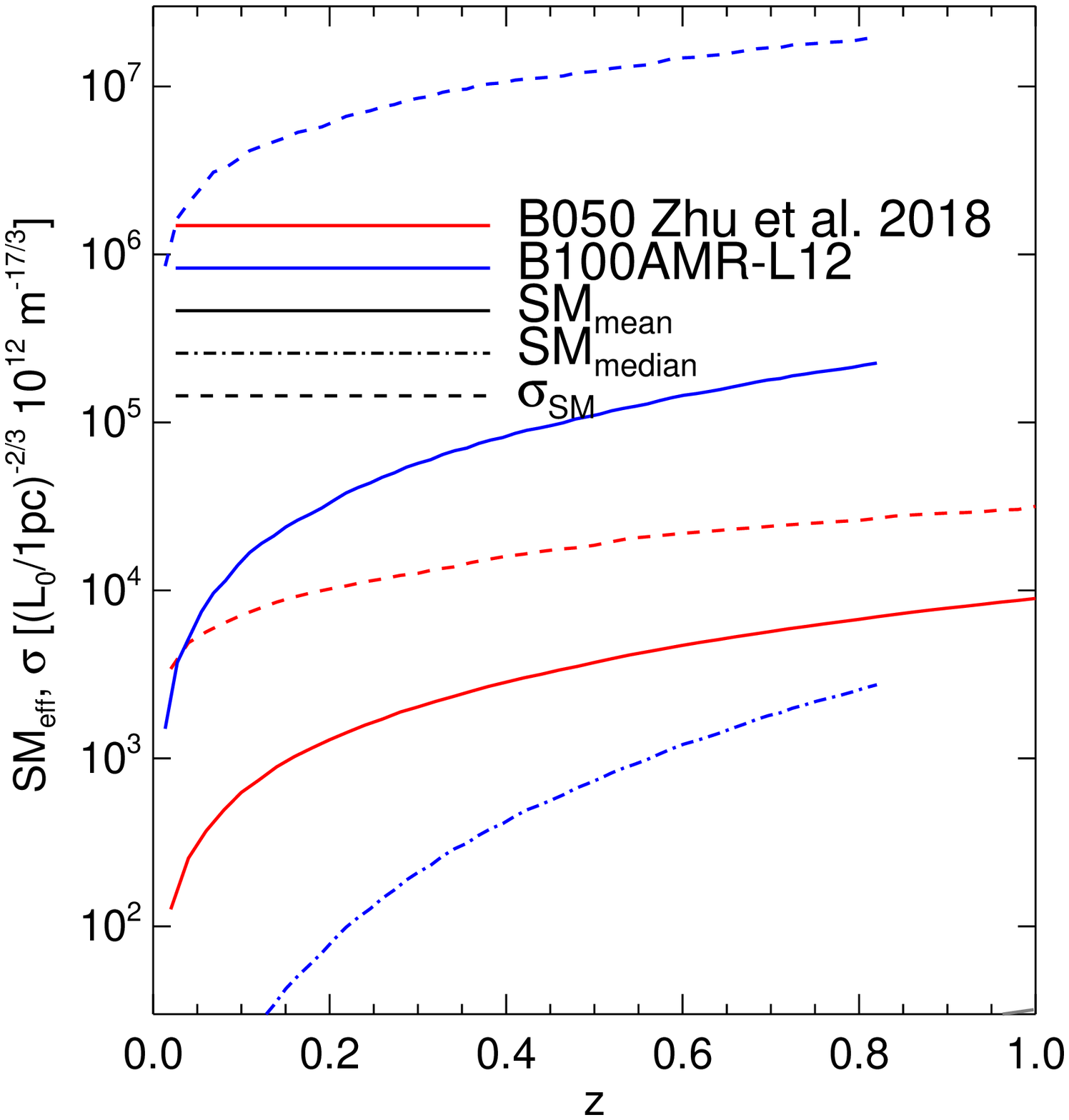}
\includegraphics[width=0.88\columnwidth,trim=0 0 0 15,clip]{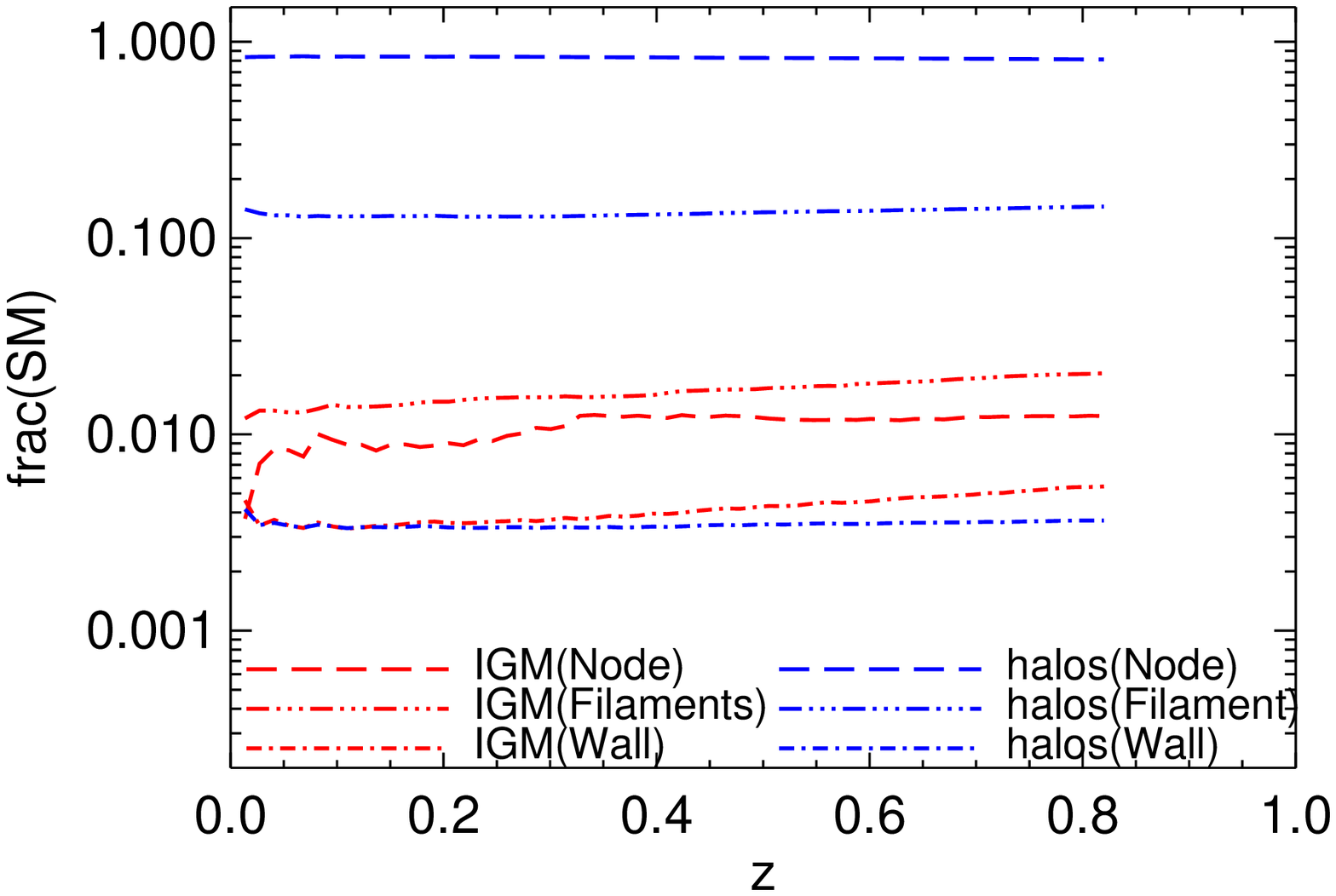}
\caption{Top: Solid, dotted-dashed and dashed lines in blue color indicate the mean, median and standard deviation of $\rm{SM_{IGM}(z)+SM_{halos}(z)}$, based on the AMR simulation in this work. Red lines are corresponding results in Z18. Bottom: Same as Fig.~\ref{fig:dm_frac}, but for $\rm{SM_{eff}}$.}
\end{center}
\label{fig:sm_z_frac}
\end{figure}

\begin{figure*}[htbp]
\begin{center}
\hspace{-0.0cm}
\includegraphics[angle=90,width=1.00\textwidth]{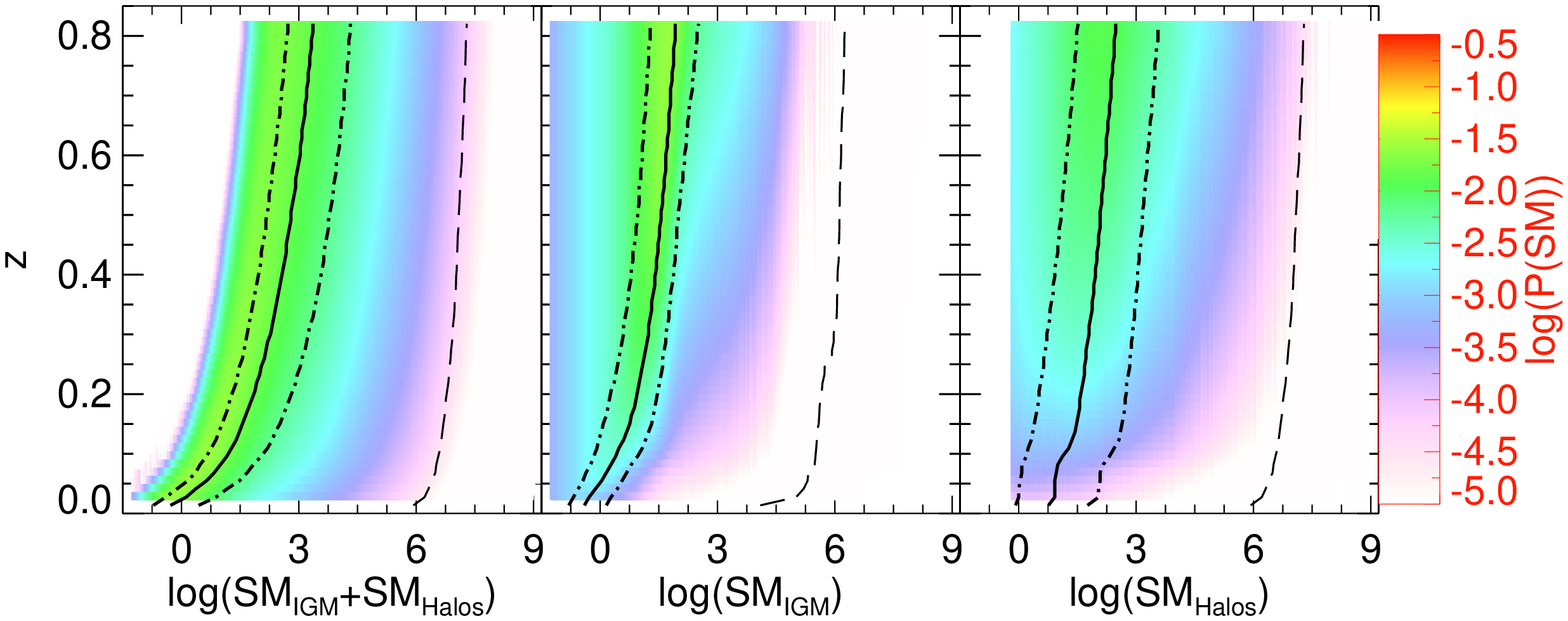}
\caption{Same as Fig.~\ref{fig:dm_pdf}, but for $\rm{SM_{eff}}$.}
\end{center}
\label{fig:sm_pdf}
\end{figure*}

\subsection{SM of the IGM and foreground halos}
Fig.~\ref{fig:sm_z_frac} presents the scattering measure caused by the IGM and foreground halos, $\rm{DM_{IGM}+DM_{halos}}$, as a function of redshift. Not surprisingly, the increased resolution in our AMR simulation in comparison with Z18 has enhanced the effective scattering measure dramatically. This trend is consistent with the results of simulations with different resolution in Z18. As demonstrated in Z18, the effective scattering measure is very sensitive to grid cells with very large over-density. Simulations with finer grids can resolve the density field in highly nonlinear regime better. 

The bottom panel of Fig.~\ref{fig:sm_z_frac} shows that the halos residing in nodes and filaments contribute about $\sim 80\%$ and $\sim 15\%$ of $\rm{SM_{IGM}+SM_{halos}}$. In contrast, the contribution from the IGM in nodes and filaments are merely around $1\%$. The IGM and foreground halos in the wall and void regions make very little contribution to the scattering. In Z18, we found that the medium in nodes and filaments contribute about $\sim 65-80\%$ and $\sim 20-30\%$ of the effective SM. Here, our investigation further indicates that the gaseous halos in nodes and filaments play dominant role in the scattering of FRBs signals throughout the path between the host halos and the Milky Way. This result is consistent with the theoretical investigation in \cite{2013ApJ...776..125M}. The standard deviation of $\rm{SM_{eff}}$ is larger than the mean value of $\rm{SM_{eff}}$ by almost two dex. Again, foreground halos is the primary factor that results in tremendous difference between different line-of-sights, as shown by Fig.~\ref{fig:sm_pdf}. Line-of-sights that have passed through the most inner region of halos can significantly enhance the standard deviation.

\subsection{DM and SM of host halos}
\begin{figure}[htbp]
\begin{center}
\includegraphics[width=0.90\columnwidth]{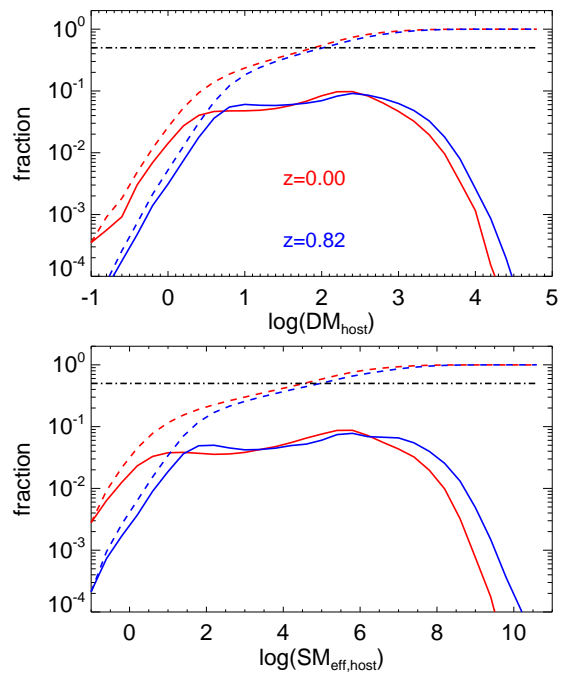}
\caption{Top: Solid lines indicate the distribution of dispersion measure, in unit of ${\rm{pc\, cm^{-3}}}$, caused by halos more massive than $1.2 \times 10^{11} M_{\odot}$ at redshift $z=0.0$(red) and $z=0.82$(blue). Dashed lines show the corresponding cumulative distribution functions. Bottom: Same as the top panel, but for the effective scattering measure, in unit of $10^{12} (L_0/\rm{1pc})^{-2/3} m^{-17/3}$, where $L_0$ is the outer scale of turbulence.}
\end{center}
\label{fig:dm_sm_host_pdf}
\end{figure}

Contribution from the medium in the hosting halos is also an important component of the total DM and SM of FRB events. However, it remains very difficult to estimate those values for a particular event at the present moment. It is usually assumed to follow a log-normal distribution for disk galaxy, based on results of previous theoretical study(e.g. \citealt{2015RAA....15.1629X}). We tackle this issue statistically as follows. We calculate the DM and effective SM along each of the 40 randomly radial trajectories for each halo with $M_h>1.2 \times 10^{11}M_{\odot}$ at certain redshifts snapshots, using Eqn.~\ref{eq:dm_hosts} and Eqn.~\ref{eq:smeff_hosts}. The number of halos over $1.2 \times 10^{11}M_{\odot}$ slightly decrease from $\sim 39500$ at $z=0.82$ to $\sim 34700$ at $z=0$. Thus, we have about 140000-160000 radial trajectories within host halos for each simulation snapshot. Here, we assume the halos in a particular snapshot have the same redshift. The red solid lines in Fig.~\ref{fig:dm_sm_host_pdf} shows the distribution of DM and effective SM associated to these trajectories at $z=0$. 

Significant variations can be found among different radial trajectories of different halos, which should be due to the inhomogeneity of the gas distribution in host halos. At $z=0.0$, $\rm{DM_{host}}$ spans a wide range of values $\sim 1-3000 \  {\rm{pc\, cm^{-3}}}$ for most of the radial trajectories, and its distribution is obviously deviated from a log-normal one, but exhibits an almost even distribution in the range $\sim 3-30 \  {\rm{pc\, cm^{-3}}}$, and then continued with a bump peaked at $\sim 300 \  {\rm{pc\, cm^{-3}}}$. The median value is $\sim 100\ {\rm{pc\, cm^{-3}}}$, in agreement with previous theoretical models( e.g., \citealt{2015RAA....15.1629X}; \citealt{2018arXiv180401548W}; \citealt{2018MNRAS.481.2320L}). However, the distribution of $\rm{DM_{host}}$ extracted from our simulation displays a significant difference from a simple log-normal form that is usually assumed in theoretical models. 

This discrepancy may result from the following factors. In our simulation study, radial trajectories can be associated with host halos of various masses, morphology types and randomly selected view angles. In the previous theoretical study based on scaled models of smooth electron distribution in galaxy(e.g. \citealt{2015RAA....15.1629X}), the distribution of $\rm{DM_{host}}$ for a galaxy with particular stellar mass, morphology type and view angle can be fitted by a log-normal function. But the fitting parameters of the log-normal function will vary notably if either the galaxy mass or galaxy type as well as view angle is changed. Moreover, as shown in Figure 1, the gas distribution in halos is not smooth and contains high density clumps, which may contribute to this discrepancy.

The scattering measure of hosts exhibits a similar distribution to $\rm{DM_{host}}$ in the range $\rm{SM_{eff,host}}$ $\sim 1-10^8 \times 10^{12}(L_0/\rm{1pc})^{-2/3}  m^{-17/3}$, with a median value $\sim10^5 \times 10^{12}(L_0/\rm{1pc})^{-2/3}  m^{-17/3}$. The results of DM and SM caused by hosts at $z=0.82$, shown with blue lines in Fig.~\ref{fig:dm_sm_host_pdf}, are basically similar to that at $z=0.0$, but with larger median value of DM and SM. This is not surprising, actually, we selected halos from the same simulation, and the evolution of those halos is relatively slow at low redshifts. On the other hand, higher physical density at high redshifts should have shifted the overall distribution toward larger dispersion and scattering measure. Note that, all the radial propagation paths within the hosts in our calculation are assumed to start from the halo center, neglecting the spatial distribution of FRB events. It may mildly overestimate both the contribution to DM and SM by host halos. 
Since our simulation has a limited spatial resolution in halos, further investigation with higher-resolution simulations is required to resolve finer structure of host halos and give more reliable results.

\section{the DM-$z$ and DM-$\tau$ relation}
In this section, we investigate the DM distribution, DM-redshift relation and the DM-$\tau$ relation of FRB events, and quantify the relative importance of the host halos, foreground halos and the IGM on DM and $\tau$ of FRB events. So far, more than 100 FRB events have been reported. In our study, we use the information of 124 source available in the literature \footnote{\url{http://www.frbcat.org}} (\citealt{2016PASA...33...45P}) by the end of May, 2020. Among them, 38 events have reported values of time broadening scale $\tau$, and another 36 events have reported upper limits on $\tau$. For repeating FRB sources, the lowest value of scattering time in the literature is used. 

 We produce 50000 mock sources with a uniform redshift distribution in the range $z\sim 0.0-0.82$. Note that, we place a redshift cutoff at $z=0.82$ by hand due to the limited snapshots generated by our cosmological simulation. Continuous distribution of the IGM and foreground halos are only available at $z \leq 0.82$. In the reality, there should be FRB events occurred at redshifts higher than 0.82. We will discuss the limitation later. Each mock source is randomly associated with one of the 100000 lines-of-sights sampling for the IGM and foreground halos as described in the last section, and with one of the radial trajectories associated to halos more massive than $1.2 \times 10^{11}M_{\odot}$ in the snapshots corresponding to FRBs redshift. The gas density, dispersion and scatter measure along the l.o.s, start from z=0 and end at the sources redshift, and along the radial trajectory to each mock source are used to calculate its total DM and scattering time $\tau$, under certain assumptions about the turbulence in different medium components. 
 
 For an individual gas cell along the light path to each mock source, we can identify which components it belongs to, i.e., either the IGM, or the foreground halos, or the host halo. Therefore, we can separate the contributions to the total extragalactic DM and $\tau$ of any mock sources from the three types of medium. In the following study, mock sources having a total extragalactic DM larger than $\rm{3500\, pc\, cm^{-3}}$ will not be taken into accounted, because the highest DM of observed events is around 2600 $\rm{pc\, cm^{-3}}$. In result, about 3 percents of the mock sources are excluded.  As we will show, the distribution of extragalactic DM and $\tau$ of mock sources are generally in agreement with the observed events.

\begin{figure*}[htbp]
\begin{center}
\hspace{-0.5cm}
\includegraphics[width=0.68\columnwidth, trim=0 0 0 0,clip]{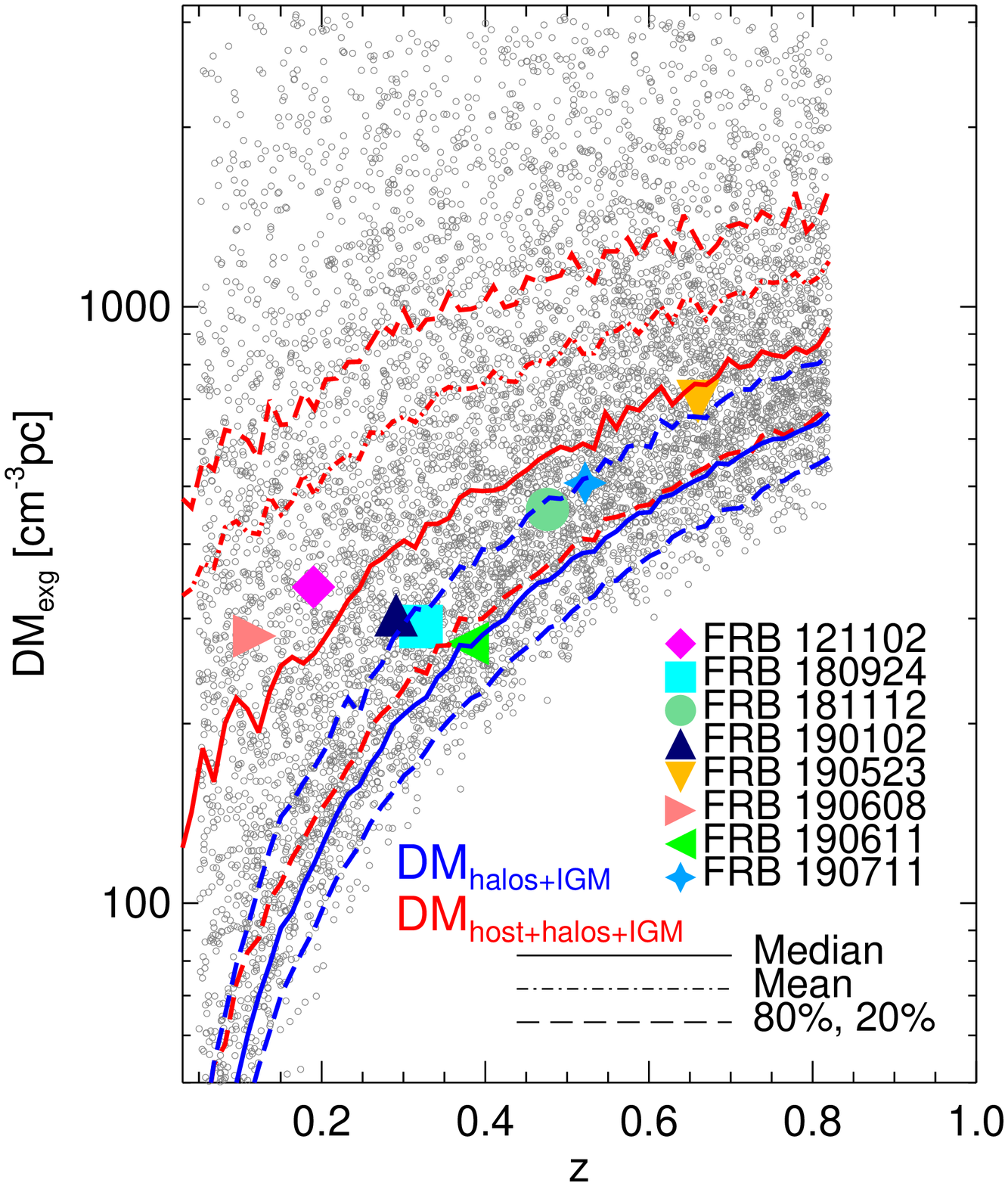}
\includegraphics[width=0.68\columnwidth, trim=0 0 0 0,clip]{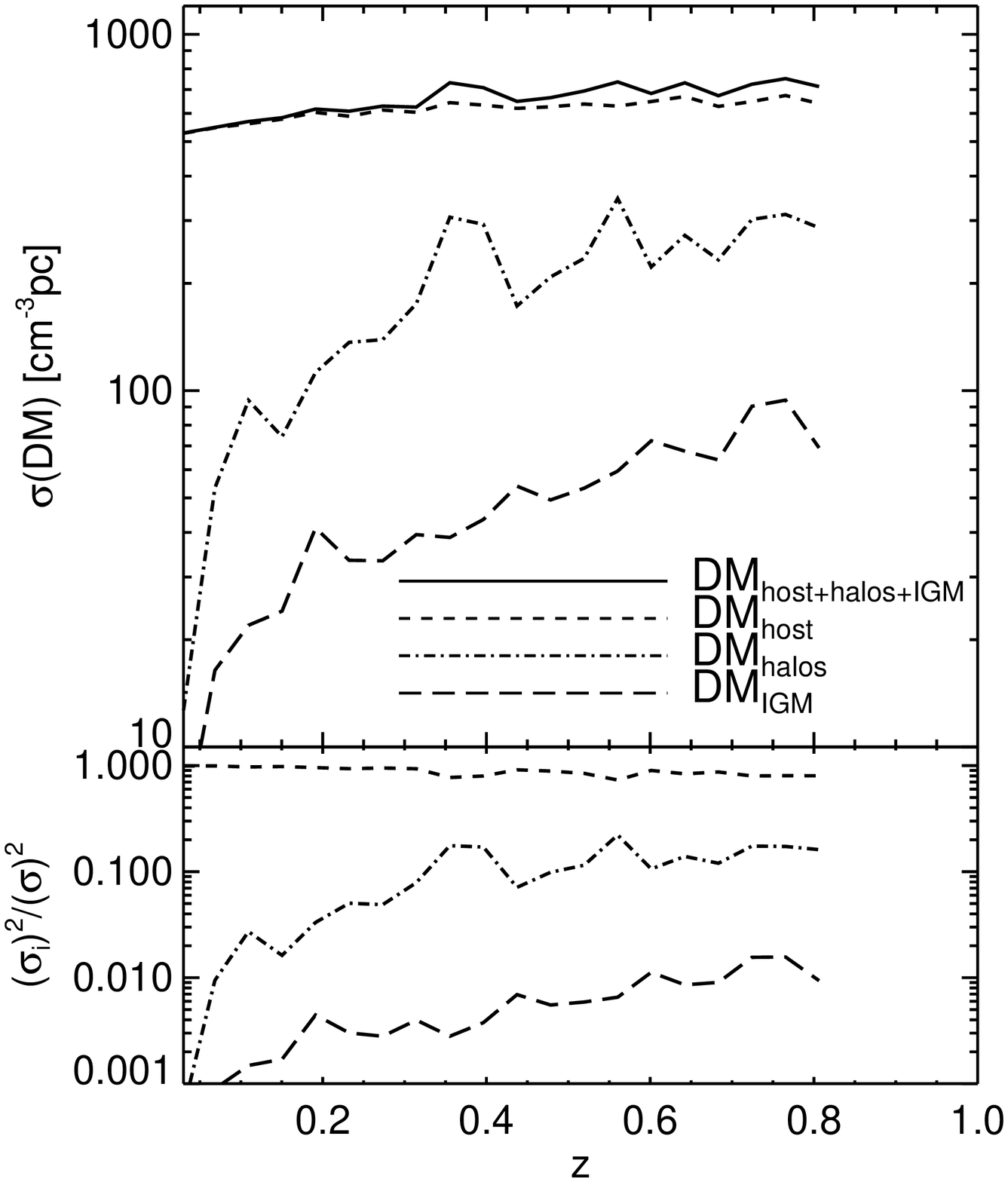}
\includegraphics[width=0.68\columnwidth, trim=0 0 0 0,clip]{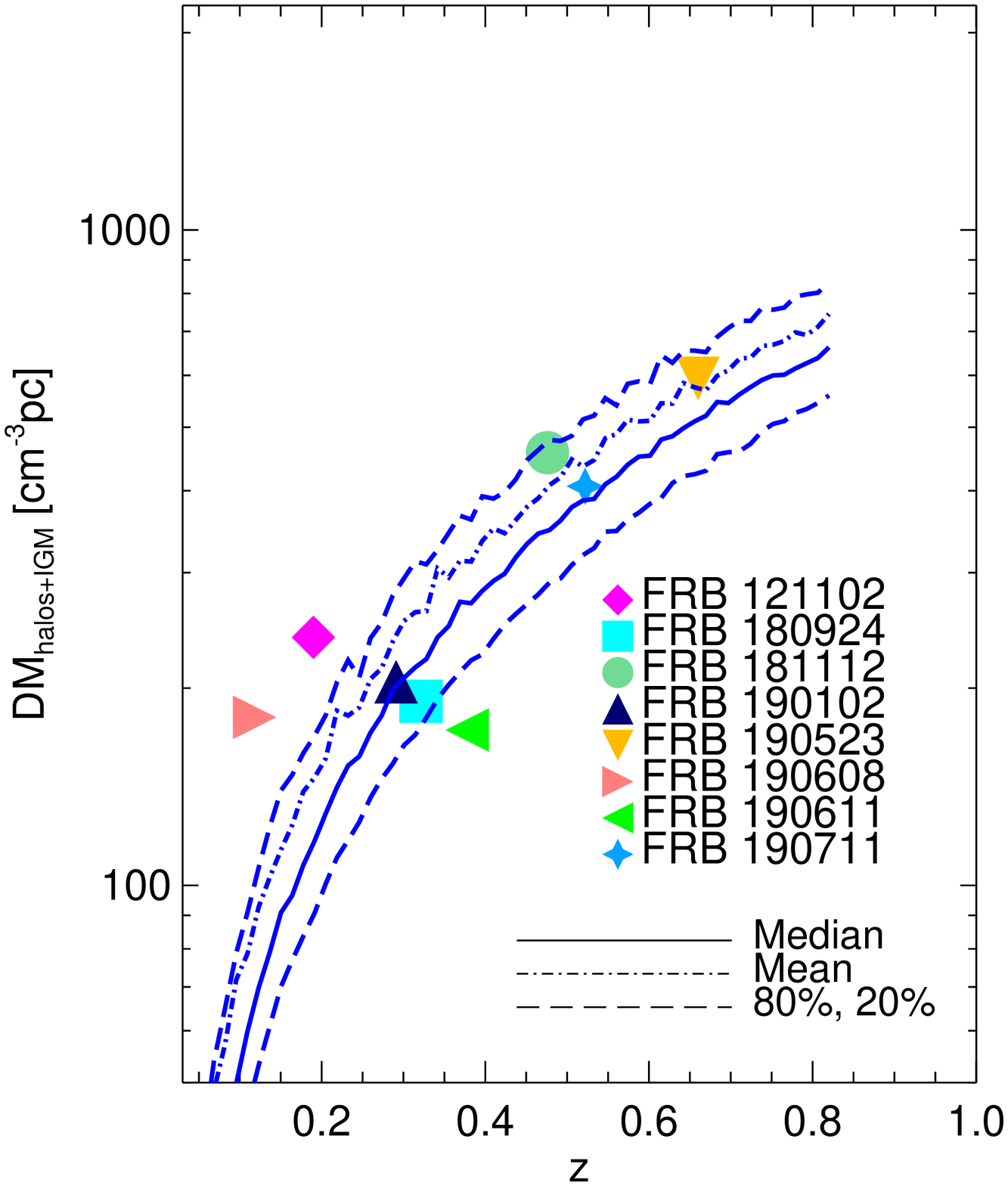}
\caption{Left: Redshift vs. $\rm{DM_{exg}}$ scatter plot of the mock sources, indicated by gray open circles. Lines indicate the median(solid), mean(dotted-dashed) , 80$\%$ and 20$\%$(long dashed) percentage of $\rm{DM_{host}+DM_{halos}+DM_{IGM}}$(red) and $\rm{DM_{halos}+DM_{IGM}}$(blue) of mock sources. Different filled symbols are the 8 FRB events whose host galaxies have been localized. Middle: Variances in the $\rm{DM_{exg}}$ of mock sources, and the contributions from different components of intervening medium. Right: $\rm{DM_{halos}+DM_{IGM}}$ of 8 localized FRB events, assuming $\rm{DM_{host}=100 \, cm^{-3} pc}$. Blue lines indicate the statistics of our mock sources. FRB 121102, 180924, 181112, 190102, 190523, 190608, 190611, 190711, listed from top to bottom in the legend, are displayed with symbols in colors of magenta, cyan, aquamarine, navy, gold, pink, green and sky respectively.}
\end{center}
\label{fig:dm_zred_var}
\end{figure*}

\begin{figure*}[htbp]
\begin{center}
\includegraphics[width=0.95\columnwidth, trim=0 0 0 0,clip]{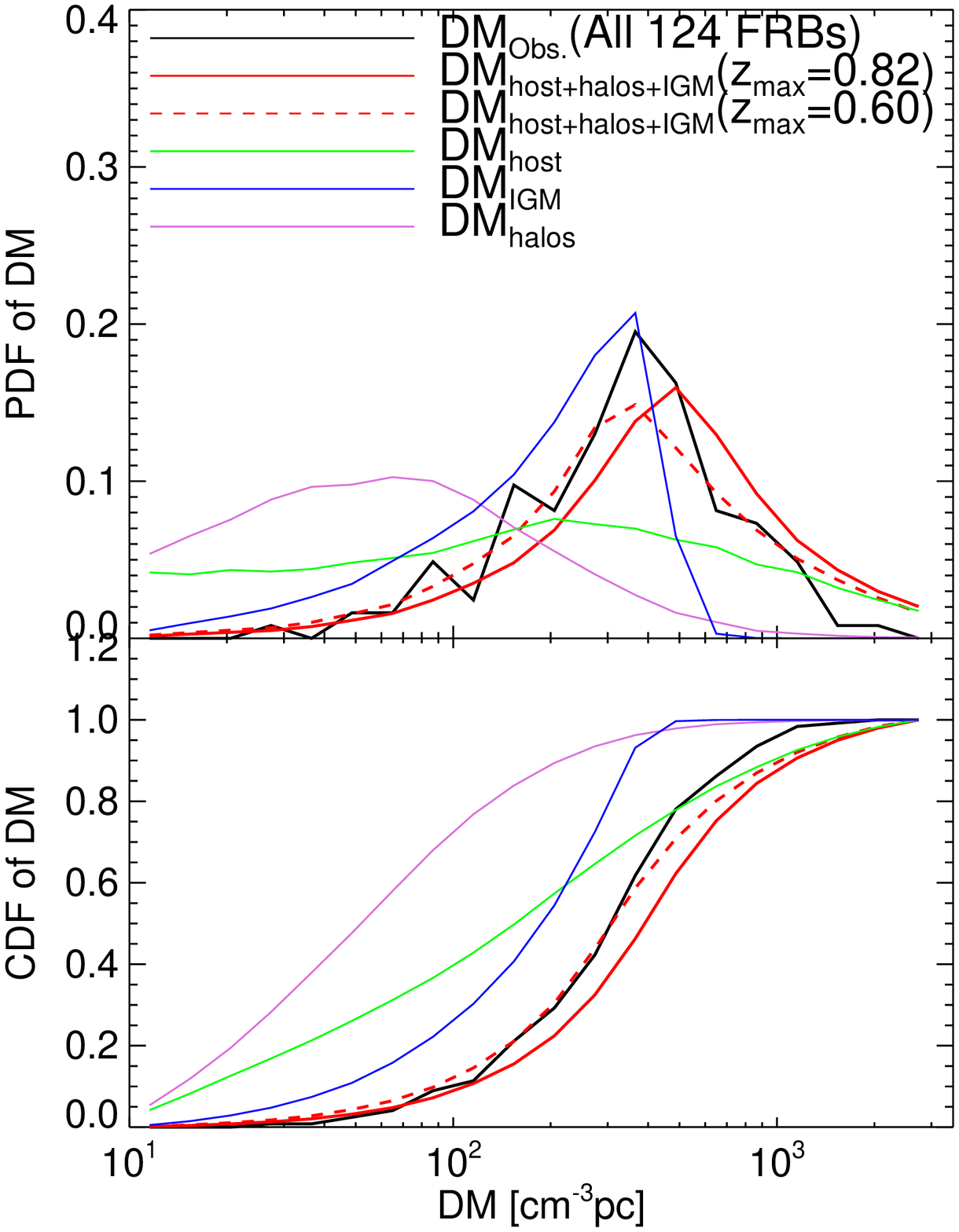}
\includegraphics[width=0.95\columnwidth, trim=0 0 0 0,clip]{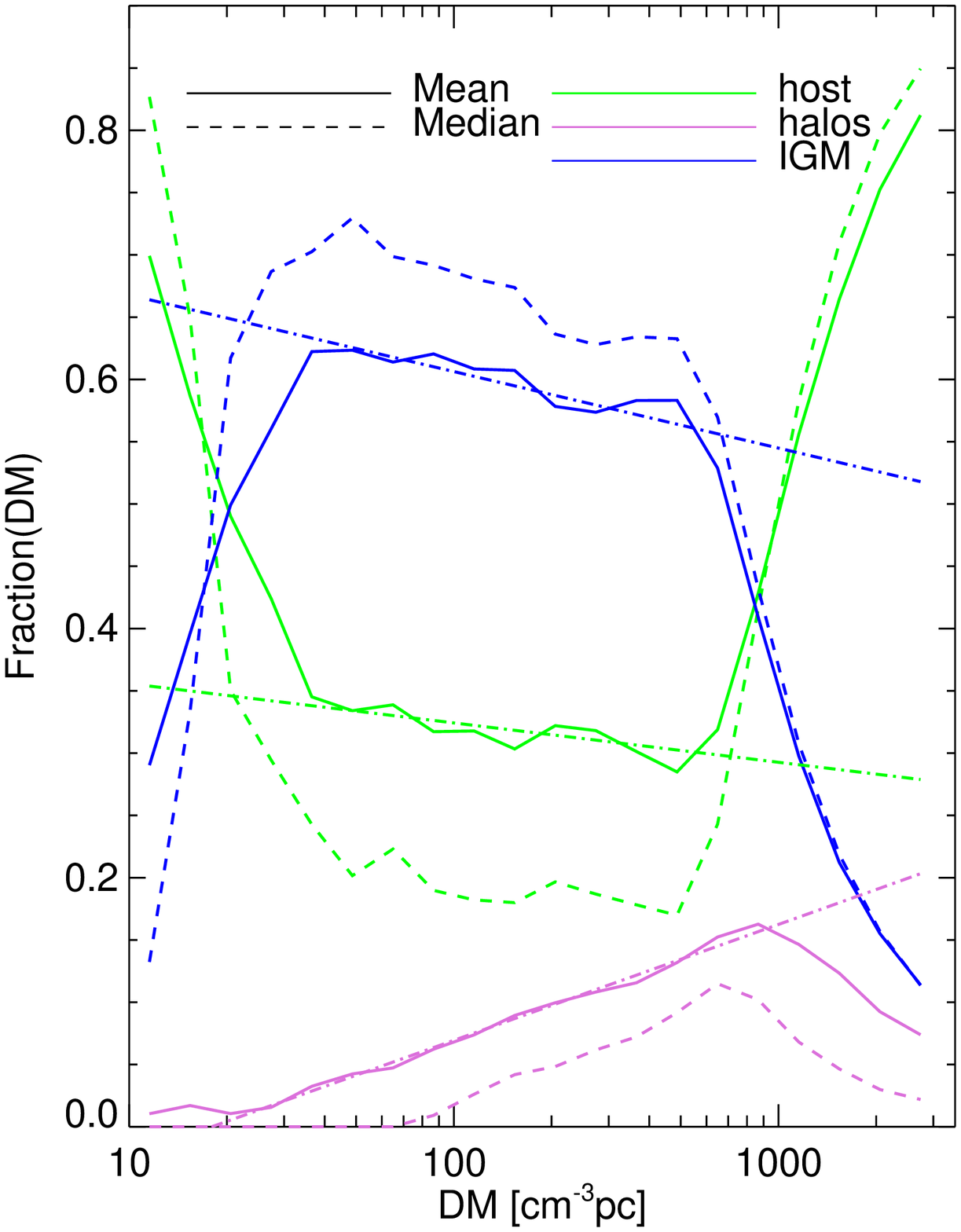}
\caption{Left: The probability distribution function(top) and cumulative distribution function(bottom) of the total extragalactic DM of mock sources, against with the observed sources(black solid line). Right: The mean(solid) and median(dashed) fractions of $\rm{DM_{exg}}$ of mock sources caused by the IGM(blue), foreground halos(purple) and host halos(green) as a function of the total $\rm{DM_{exg}}$.}
\end{center}
\label{fig:dm_pdf_cont}
\end{figure*}

\subsection{DM-z relation, DM distribution}

With the mock FRB sources. we give a scatter plot of the total extragalactic DM, i.e., $\rm{DM_{exg}=DM_{host+halos+IGM}}$ $=
\rm{DM_{host}+DM_{halos}+DM_{IGM}}$, against theirs redshifts in the left panel of Fig.~\ref{fig:dm_zred_var}. For a given redshift, the total DM displays a significant variations. We further measure the variances caused by different components and present the result in the middle panel of Fig.~\ref{fig:dm_zred_var}. Clearly, the host halo overwhelmingly dominates the variance in the total DM, $\sigma^2(\rm{DM_{exg}})$, accounting for nearly $99\%$ at $z<0.2$, and the fraction remains to be around $\sim 90\%$ till $z=0.8$.  The contribution to $\sigma^2(\rm{DM_{exg}})$ from foreground halos amounts only $\sim 1\%$ at $z<0.2$, and increases gradually to $\sim 10\%$ at $z>0.4$. In addition, the IGM makes a contribution about one tenth of that from the foreground halos to $\sigma^2(\rm{DM_{exg}})$.

To date, the redshifts of eight localized FRB events have been available (\citealt{2017ApJ...834L...7T}; \citealt{2019Sci...365..565B}; \citealt{2019Natur.572..352R}; \citealt{2020Natur.577..190M}; \citealt{2020Natur.581..391M}). We plot them with filled symbols in the left panel of Fig.~\ref{fig:dm_zred_var}, assuming that the contribution from the Milky Way's halo is $30 \rm{cm^{-3} pc}$ following \citealt{2016arXiv160505890C}. Except for FRB 190611, most of them show minor or moderate deviations, $\delta z\lesssim 0.1$, from the median DM-z relation of our mock sources. Therefore, for those observed FRB events without any information of host, the median DM-z relation of our mock sources allow us to give a crude estimation of their redshifts. The fitting result for the median DM as a function of redshift is as follows,
\begin{equation}
\label{eq:dm_all_z_fit}
\begin{aligned}
\rm{log_{10}DM_{exg}(z)}=3.04+1.00\rm{log}_{10}(z)+0.25(\rm{log}_{10}(z))^2. 
\end{aligned}
\end{equation}

Meanwhile, the median DM caused by the foreground gaseous halos and the IGM of mock sources fulfills the fitting formula of Eqn.~\ref{eq:dm_z_fit}. If we assign the dispersion induced by host halos the same value as the median value obtained in section 3.3, i.e. $\rm{DM_{host}=100 \, cm^{-3} pc}$, the redshift of observed events can be estimated using Eqn.~\ref{eq:dm_z_fit}. The right panel of Fig.~\ref{fig:dm_zred_var} demonstrated that the redshifts of 5 localized FRB events can be recovered with error $\delta z \lesssim 0.05$ following this scheme. Yet, the errors for other 3 localized events are about $\delta z \sim 0.15$. The uncertainty in the dispersion measure caused by the host should be the primary source of such errors. For instance, the redshift of the event FRB 121102 is predicted to be $\sim 0.25$ and $\sim 0.35$ according to Eqn.~\ref{eq:dm_all_z_fit} and Eqn.~\ref{eq:dm_z_fit} respectively, which are both higher than the real value of $z=0.1927$(\citealt{2017ApJ...834L...7T}). In comparison, if the contribution from the host is not taken into account, the redshift of this event estimated from the DM can be as high as 0.68(e.g. \citealt{2019ApJ...886..135P}). The uncertainty in the dispersion measure arisen from foreground halos would be the secondary source of errors in estimating redshift from $\rm{DM_{exg}}$.

The left panel in Fig.~\ref{fig:dm_pdf_cont} compares the probability density function(PDF) and cumulative density functions(CDF) of the total extragalactic DM of mock sources with the observed events, and indicates a good agreement between the simulation and observations, though the mock sample has a slight higher fraction of sources with larger values of DM$>600 {\rm{pc\, cm^{-3}}}$. This excess can partly be attributed to mock samples with larger $\rm{DM_{host}}$. It suggests that a number of FRBs events at redshift $z<1.0$ but with larger $\rm{DM}$ may have been missed out by current observations. Alternatively, the discrepancy can be alleviated by placing a lower value of maximum redshifts $z_{max}$ of mock sources. Currently, the highest redshift of FRB events is still unclear. Due to limited number of stored snapshots, the maximum redshift of our mock sources has been set up to $0.82$. In comparison, we plot the results with $z_{max}=0.6$ in Fig.~\ref{fig:dm_pdf_cont} , which can provide better agreement with the observation.  

The relative importance to the extragalactic DM from different components of intervening medium is crucial for the understanding of FRB, and its application to probe the cosmic baryons. We carry out an overall analysis on the relative importance in our mock samples. Fig.~\ref{fig:dm_pdf_cont} shows the mean and median fractions of DM contributed by the three types of medium as a function of the total extragalactic DM. For sources with $\rm{DM_{exg}}$ either less than 20 or larger than $800 {\rm{pc\, cm^{-3}}}$, the host halos make a dominant contribution to $\rm{DM_{exg}}$. For these sources, the secondary contribution to $\rm{DM_{exg}}$ arises from the IGM. While for sources with $\rm{DM_{exg}}$ in the range of $20-800 {\rm{pc\, cm^{-3}}}$, the primary contributor is the IGM, followed by the host halos and foreground halos in turn. The mean fraction of $\rm{DM_{exg}}$ caused by the IGM declines monotonous from $\sim 60\%$ at $\rm{DM_{exg}}=20 {\rm{pc\, cm^{-3}}}$ to $\sim 50\%$ at $\rm{DM_{exg}}=800 {\rm{pc\, cm^{-3}}}$. Simultaneously, the mean contribution from the host halos drops from $35\%$ to $30\%$. 

It should be noted that, the transition of relative importance between the IGM and host halos at $\rm{DM_{exg}}\sim 800-900 {\rm{pc\, cm^{-3}}}$ should result from the maximum redshift of mock sources $z_{max}=0.82$ that placed by hand. As shown in Fig.~\ref{fig:dm_z}, the mean and median DM induced by the IGM and foreground halos are increasing with redshift and will be around $600-900 \rm{pc\, cm^{-3}}$ at $z=0.82$. 
Hence, our mock sources with $\rm{DM_{exg}} \geq 600 {\rm{pc\, cm^{-3}}}$ is inevitably biased by those with relatively large $\rm{DM}_{host}$. If the upper limit on the redshift of sources was increased, it would be expected that the behavior of fractions of DM found in the range $20-800 {\rm{pc\, cm^{-3}}}$ can be extended to higher DM ends. If we naively extrapolate the fitting curve within the range $20-800 {\rm{pc\, cm^{-3}}}$ to higher DM range linearly, the IGM would still dominate the total DM even at $\rm{DM_{exg}=5000 pc\, cm^{-3}}$. Meanwhile, the average contribution from host halos may drop below $\sim 30\%$ of the $\rm{DM_{exg}}$ and keep declining slowly as $\rm{DM_{exg}}$ increases, and would stands as the secondary contributor till $\rm{DM_{exg} \sim 2000-3000 pc\,cm^{-3}}$. However, it should be emphasised here that these fractions are obtained by taking average over a large amount of mock sources. For a single event, a striking variance is expected. In addition, the contribution from local surrounding medium is not taken into account in our models. 

\subsection{Calculation of the scattering time $\tau$}
We shall first introduce the models and calculation of the scattering time $\tau$, before describing the DM-$\tau$ relation of mock sources. Assuming that the turbulent medium can be described by the Kolmogorov turbulence model, the temporal broadening time caused by a medium extending between $z$ and $z+\Delta z$ for a source at $z_f$ is given by(e.g., \citealt{2013ApJ...776..125M})
\begin{subequations}
\begin{equation}
\begin{aligned}
\Delta \tau=3.32  \times10^{-4}(1+z)^{-1}(\frac{\lambda_0}{30\rm{cm}})^4(\frac{D_{\rm{eff}}}{1\rm{Gpc}})\\
\times(\frac{\rm{\Delta SM}_{\rm{eff}}}{10^{12}{\rm{m}}^{-17/3}})(\frac{l_0}{1 \rm{AU}})^{-1/3} \, \rm{ms}, r_{\rm{diff}} < l_0,
\end{aligned}
\end{equation}

\begin{equation}
\begin{aligned}
\Delta \tau=9.50  \times10^{-4}(1+z)^{-1}(\frac{\lambda_0}{30\rm{cm}})^{22/5}(\frac{D_{\rm{eff}}}{1\rm{Gpc}})\\
\times(\frac{\rm{\Delta SM}_{\rm{eff}}}{10^{12}{\rm{m}}^{-17/3}})^{6/5} \,\rm{ms}, r_{\rm{diff}} > l_0,\\
\end{aligned}
\end{equation}
\end{subequations}
where $\rm{{\Delta SM}_{eff}}$ is the effective scattering measure of this medium between $z$ and $z+\Delta z$, $\lambda_0$ is the observed wavelength; $D_{\rm{eff}}$ is defined as $=D_LD_{LS}/D_{S}$, with $D_L$, $D_S$, and $D_{LS}$ being the angular diameter distance of the screen of scattering medium(centered at $z+\Delta z/2$), the FRB source(at $z_f$) to us, and of the source to the screen of scattering medium respectively. For each mock source, we measure the value of $\Delta \tau$ at successive redshift intervals, using the gas density and scattering measure along the chosen LOS and trajectory toward it, and then sum up the individual $\Delta \tau$ to estimate the total scattering time $\tau$. For each gas cell along a light path, we could identify which components it belongs to, i.e., either the IGM, or the foreground halos, or the host halos. Accordingly, we can directly know the separate contribution to $\tau$ from those three components. Note that, the geometric effect due to $\rm{D_{eff}}$ is precisely calculated in this work, which is a substantial improvement with respect to Z18.  

The diffractive length scale $r_{\rm{diff}}$ for Kolmogorov turbulence with index $\beta=11/3$ is given by(\citealt{2013ApJ...776..125M}): 
\begin{subequations}
\begin{equation}
\begin{aligned}
r_{\rm{diff}} = 2.7\times 10^{10}  (\frac{\lambda_0}{30 \rm{cm}})^{-1} (\frac{\rm{SM}_{\rm{eff}}}{10^{12}{\rm{m}}^{-17/3}})^{-1/2} \\ \times (\frac{\mathit{l}_0}{1 \rm{AU}})^{1/6} \, \rm{m}, r_{\rm{diff}}<\mathit{l}_0,
\end{aligned}
\end{equation}
\begin{equation}
\begin{aligned}
r_{\rm{diff}} = 1.6\times 10^{10}  (\frac{\lambda_0}{30 \rm{cm}})^{-6/5} (\frac{\rm{SM}_{\rm{eff}}}{10^{12}{\rm{m}}^{-17/3}})^{-3/5} \, \rm{m},\\ r_{\rm{diff}}>\mathit{l}_0.
\end{aligned}
\end{equation}
\end{subequations}
Considering the results of effective scattering measure probed in the last section, the typical diffractive length scale in the host and foreground halos and the IGM is likely to be about $\sim 10^9-10^{10}\,$m for radio signals of wavelength $\lambda \sim 30 $cm.

As shown by above equations, the scattering time scale $\tau$ depends highly on the outer and inner scale, i.e., $L_0$ and $l_0$ of turbulence, and the relative sizes of $r_{\rm{diff}}$ and $l_0$. The plausible values of the inner and outer scales of the interstellar medium in the Milky Way are around $200-1000 \rm{km}$ and $1-100$ pc, respectively (e.g., \citealt{2016arXiv160505890C}). The outer and inner scales in the host halos and foreground halos might be comparable to, or moderately larger than that in the MW. The central regions of these halos would resemble the MW to some extent, especially for those halos containing a large fraction of gas in the central region. On the other hand, more and more observations in the past decades reveal that the CGM is in a complex and multiphase state, containing sub kpc gas clumps(e.g., \citealt{2012ApJ...750...67R}; \citealt{2015MNRAS.446...18C}; \citealt{2016ApJ...833...54W}); \citealt{2017ARA&A..55..389T}). Meanwhile, recent simulations using delicate designed refinement strategy show that the circumgalactic medium exhibits significant turbulent behavior, and contains more cool dense filaments and clumps, and eddies when the resolution is boosted to a few kpc, and even sub-kpc(e.g. \citealt{2019ApJ...882..156H}; \citealt{2019MNRAS.483.4040S};  \citealt{2020arXiv200610058B}). Star formation can occurs in those dense filaments and clumps in the CGM(e.g. \citealt{2020arXiv200610058B}). Consequently, stellar evolution and feedback may drive turbulence in the CGM with an outer scale shorter than 1 kpc. 

Currently, there is barely any solid constraint on the inner and outer scales of the turbulence in the IGM. The driving scale could be tens of kpc to a few Mpc, depending on the specific energy injection mechanism(see discussions in \citealt{2014ApJ...785L..26L} and \citealt{2018ApJ...865..147Z}). But the information on the state of the IGM below tens of kpc is absent from both simulation and observation works. To simplify the problem, we assume the same value of outer scale $L_0$ in the IGM and host halos as well as foreground halos in most of our models. The reason behind this assumption is that, as demonstrated in Fig.~\ref{fig:sm_z_frac} and \cite{2018ApJ...865..147Z}, the scattering caused by the IGM mainly come from those medium residing in filaments and clusters(nodes). It is probably dominated by the clumps locating in filaments and clusters, i.e., belonging to halos less massive than $1.2 \times 10^{11}M_{\odot}$. 

\begin{deluxetable}{cccc}[htbp]
\tablenum{1}
\tablecaption{Inner and outer scales of turbulence in different models taken for calculating $\tau$\label{tab:inner_outer_scales}}
\tablewidth{0pt}
\tablehead{
\colhead{Model} & \colhead{Medium} &\colhead{Inner scale} & \colhead{Outer scale}
}
\decimalcolnumbers
\startdata
{ }& Host & 1000 km & 5.0pc \\
Model-A& Halos & 1000 km & 5.0pc\\
{ }& IGM & 1 AU & 5.0pc \\
\hline
{ }& Host & 1000 km & 10.0pc \\
Model-B& Halos & 1000 km & 10.0pc\\
{ }& IGM & 1000 km & 10.0pc \\
\hline
{ }& Host & 1 AU & 0.2pc \\
Model-C& Halos & 1 AU & 0.2pc\\
{ }& IGM & 1 AU & 0.2pc \\
\hline
{ }& Host & 1000 km & 5.0pc \\
Model-D& Halos & $10^6$ \rm{km} & 50.0pc\\
{ }& IGM & 1 AU & 500.0pc \\
\enddata
\end{deluxetable}

Here, we study the scattering time of FRB events in \textbf{four} models. For the first three models, the outer scale in the three types of intervening medium are assumed to be the same. We set the inner scales in different medium to some fixed values, and then adjust the outer scales to produce a scattering time scale comparable to the observations. In the first model, denoted as 'Model-A', the inner scale of the host halos and foreground halos is $1000$ km, and the inner scale of the IGM is $1 $AU. In the second and third models, denoted as 'Model-B' and 'Model-C', all the inner scale in three types of intervening medium are $1000$ km and $1$ AU respectively. As the real values of outer and inner scale in the foreground halos and the IGM are poorly constrained so far and the scattering time depends highly on these scales, we also include the fourth model, denoted as 'Model-D', in which these turbulence scales in the foreground halos and IGM are much larger than those in the host halos. We list the values of $l_0$ and $L_0$ in different models in Table \ref{tab:inner_outer_scales}. A inner scale $l_0=1000 $km would be much shorter than the typical $r_{\rm{diff}}$ in those three types of medium discussed here. While for a inner scale $l_0=1 $AU, it is probably longer than $r_{\rm{diff}}$. We will adopt the corresponding equation to estimate $\tau$ while using these two values of $l_0$. 

\begin{figure*}[htbp]
\begin{center}
\vspace{-0.5cm}
\subfigure[]{
\includegraphics[width=0.82\columnwidth]{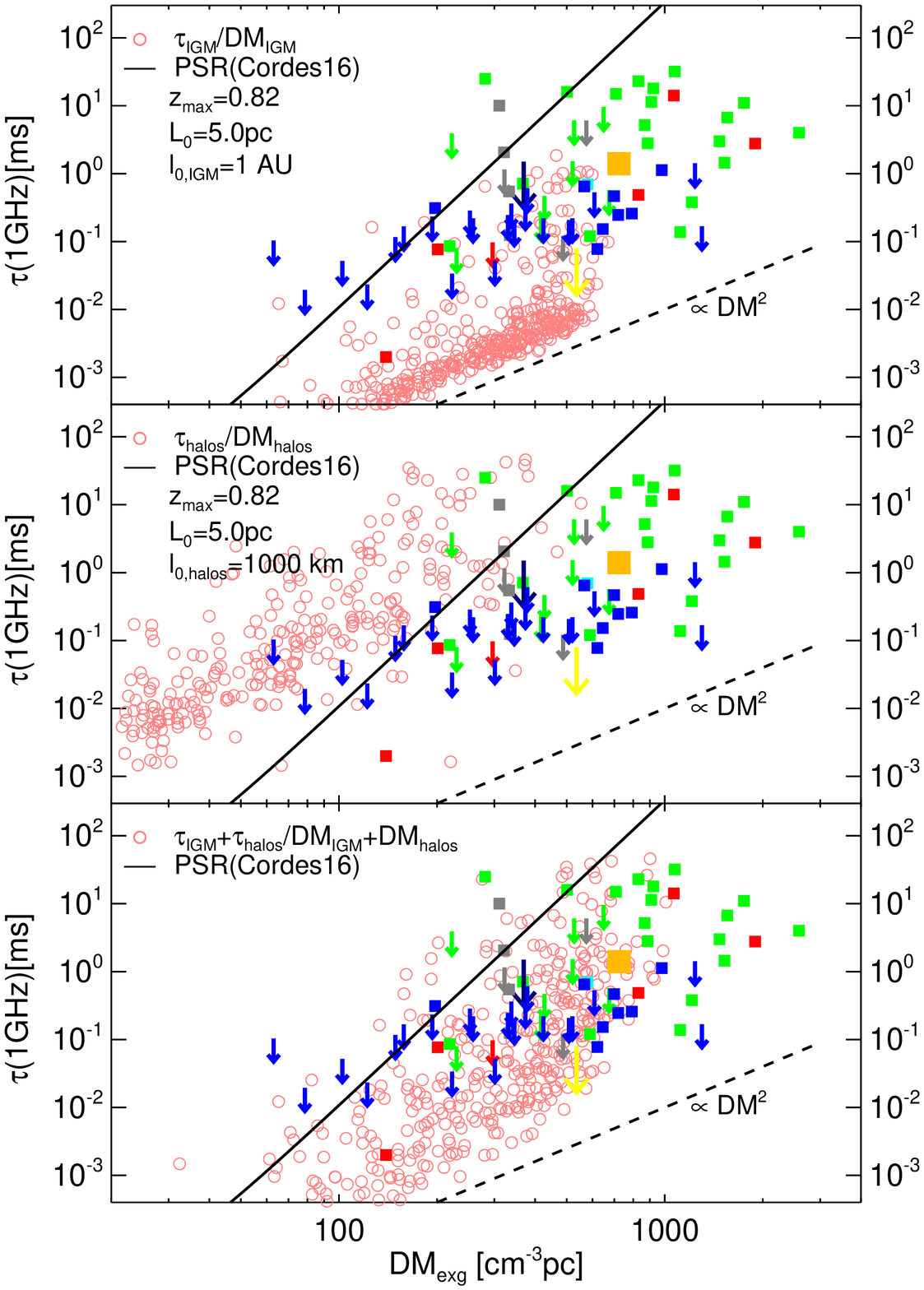}
\includegraphics[width=0.82\columnwidth]{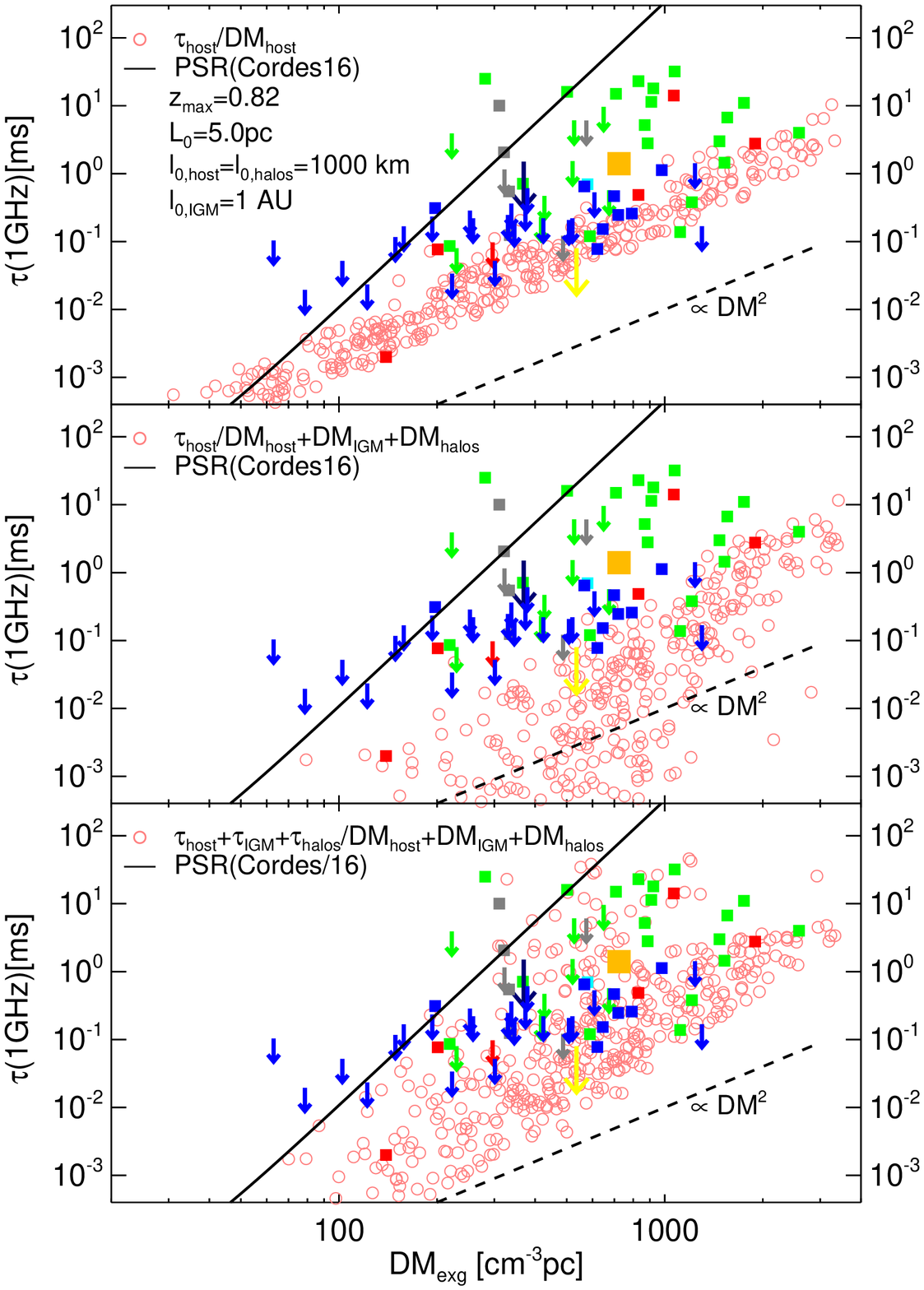}
}
\subfigure[]{
\includegraphics[width=0.82\columnwidth]{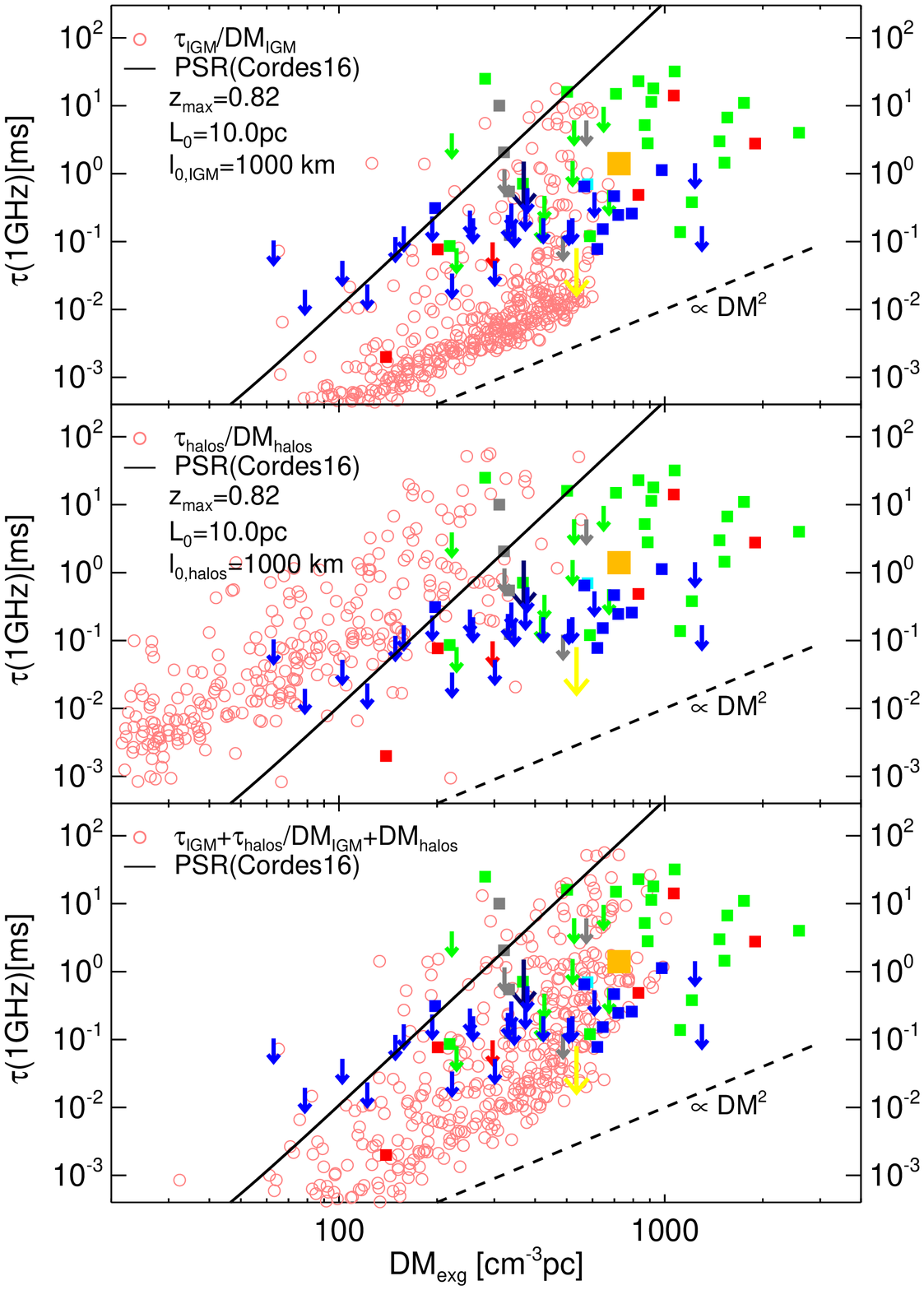}
\includegraphics[width=0.82\columnwidth]{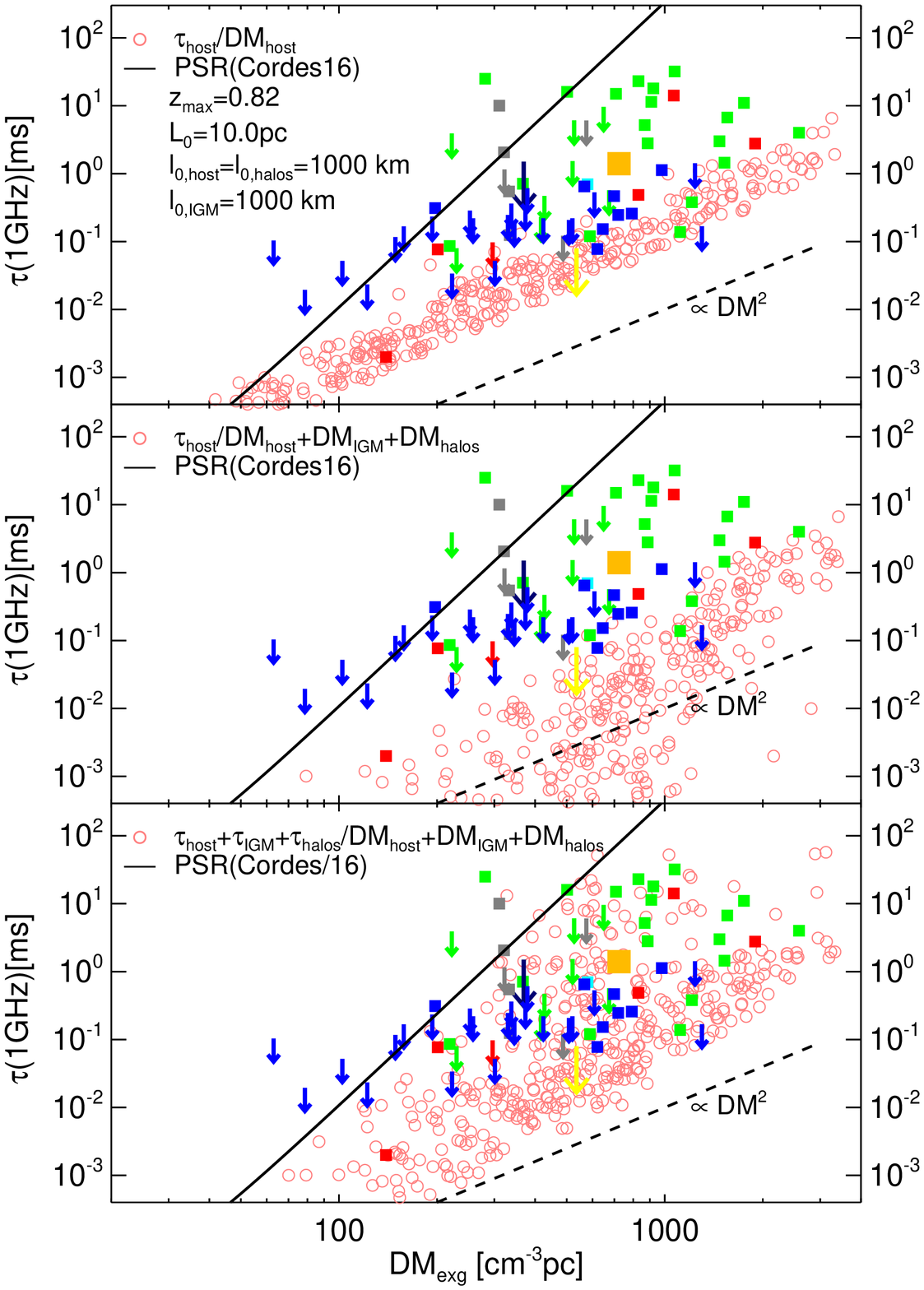}
}
\vspace{-0.2cm}
\caption{(a): The extragalactic DM-$\tau$ relation of 500 mock samples(pink circles) for Model-A(see section 4.1 for detail), against observed samples. Solid symbols(downward arrows) indicates events that have reported(upper limiter) $\tau$.  Blue, cyan, gray, green, gold, navy, red and yellow, colors indicate sources detected by CHIME, GBT, ASKAP, Parkes, DSK-10, Arecibo, UTMOST, and WSRT telescope respectively. The top-left, middle-left and top-right panels indicate that if the contribution to DM and $\tau$ is solely from the IGM, foreground halos and host halos respectively. The middle-right panel indicates that the DM of mock sources are the sum from three types of medium, while only the host halos contributes to $\tau$. The bottom-left panel shows the case if DM and $\tau$ of mock sources are caused by the IGM and foreground halos. Contributions to the DM and $\tau$ of mock sources from all of the intervening medium are accounted in the bottom-right panel. (b): results for Model-B.  }
\end{center}
\label{fig:dm_tau_modela}
\end{figure*}

\begin{figure*}[htbp]
\begin{center}
\vspace{-0.5cm}
\subfigure[]{
\includegraphics[width=0.82\columnwidth]{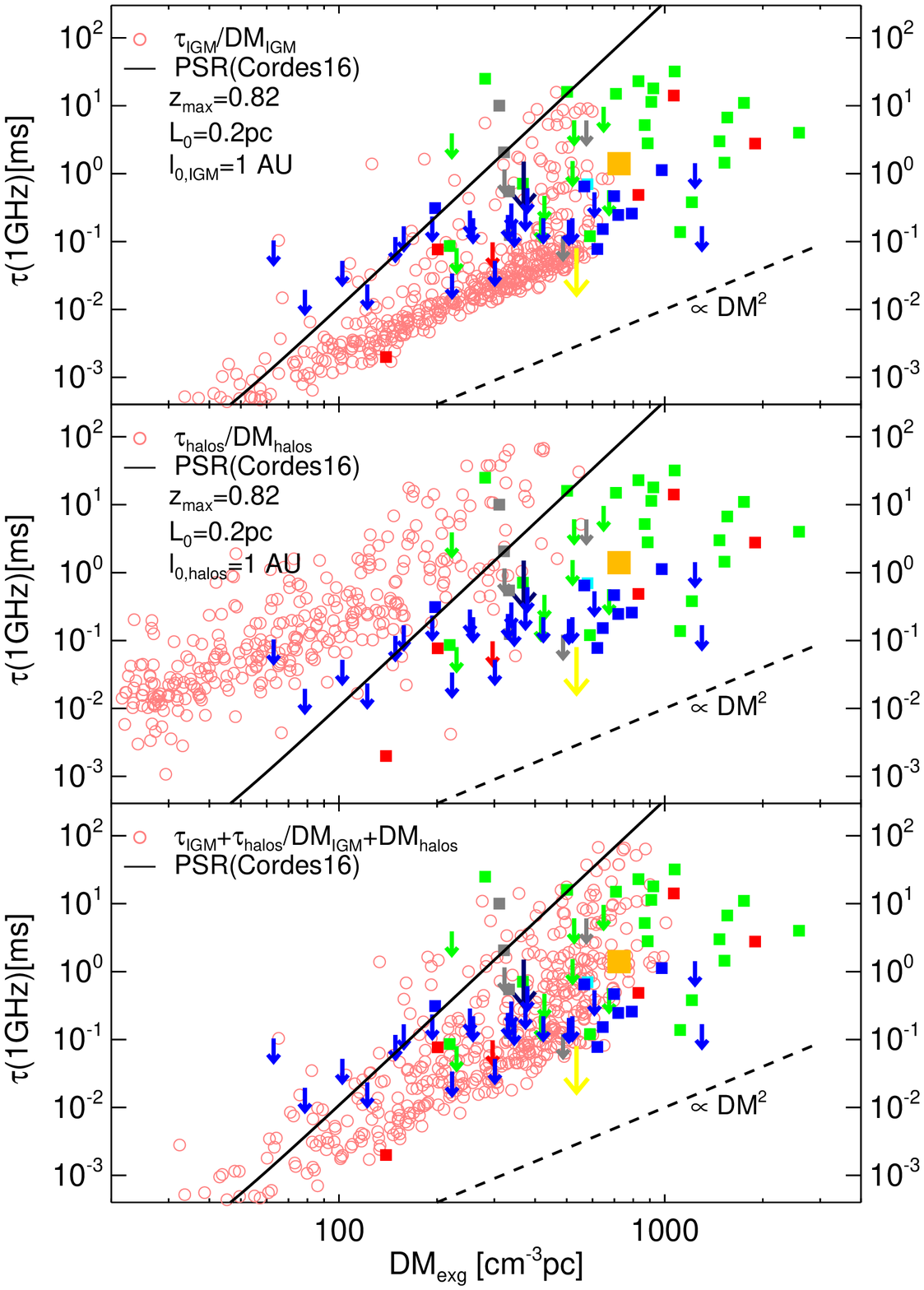}
\includegraphics[width=0.82\columnwidth]{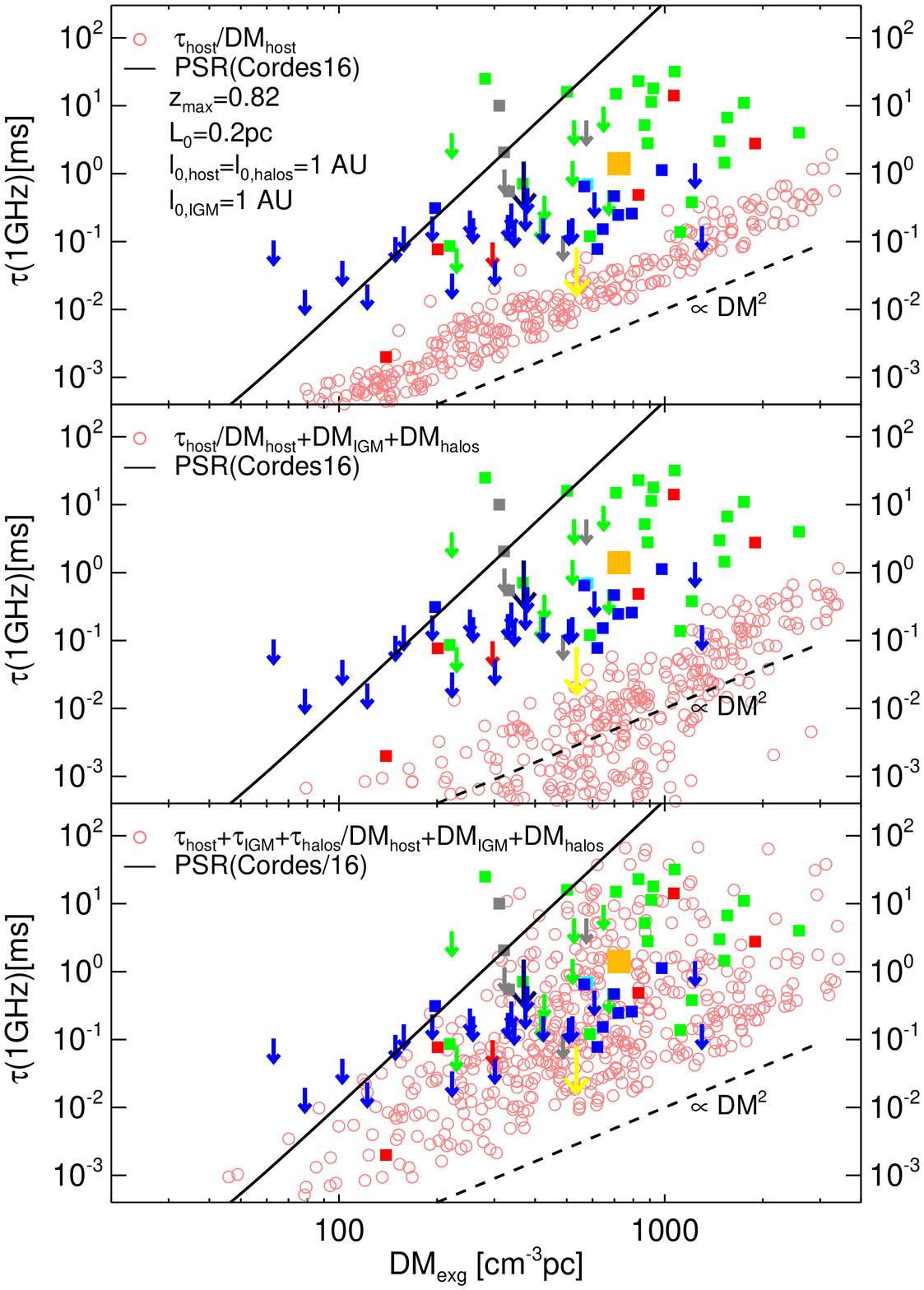}
}
\subfigure[]{
\includegraphics[width=0.82\columnwidth]{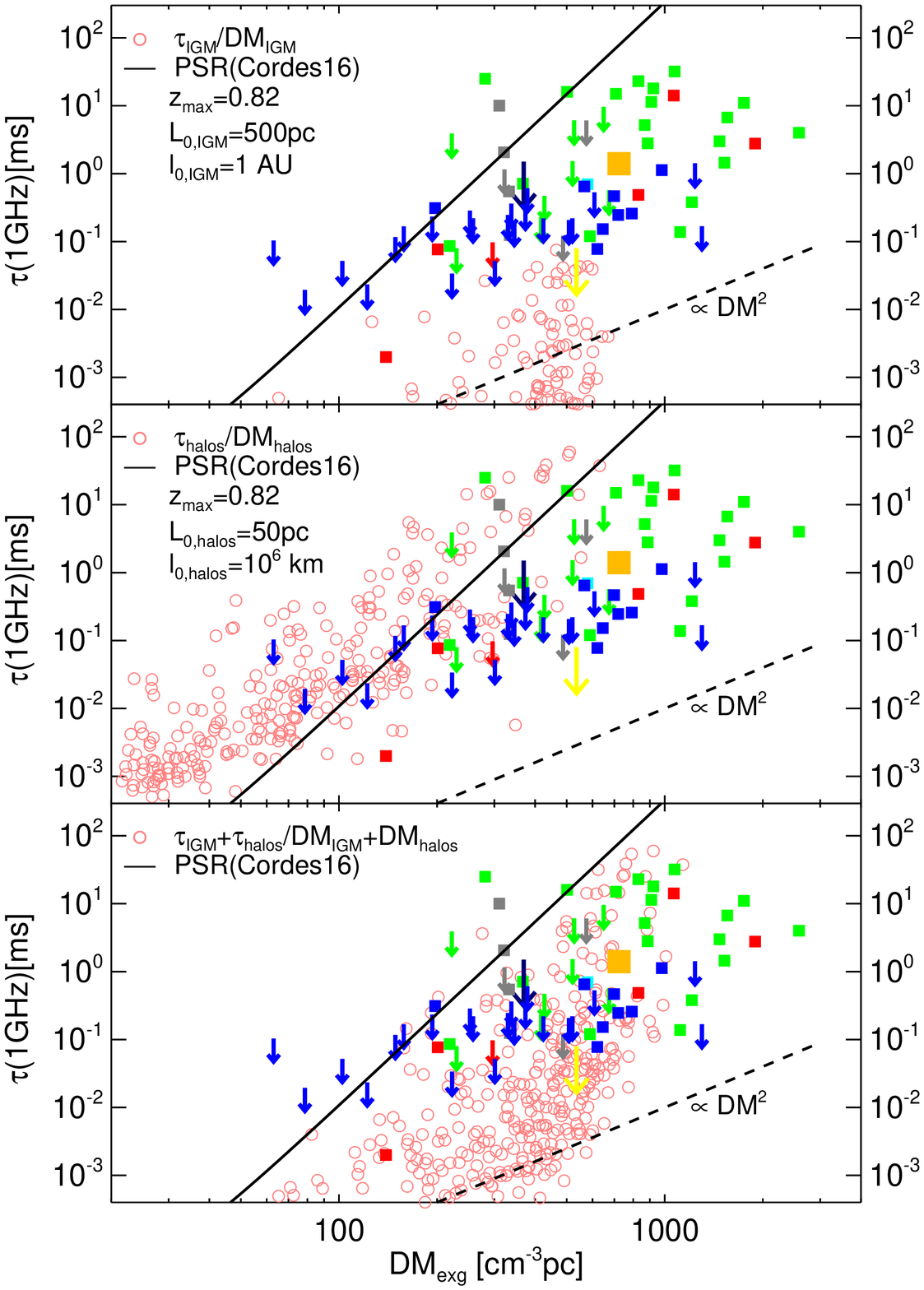}
\includegraphics[width=0.82\columnwidth]{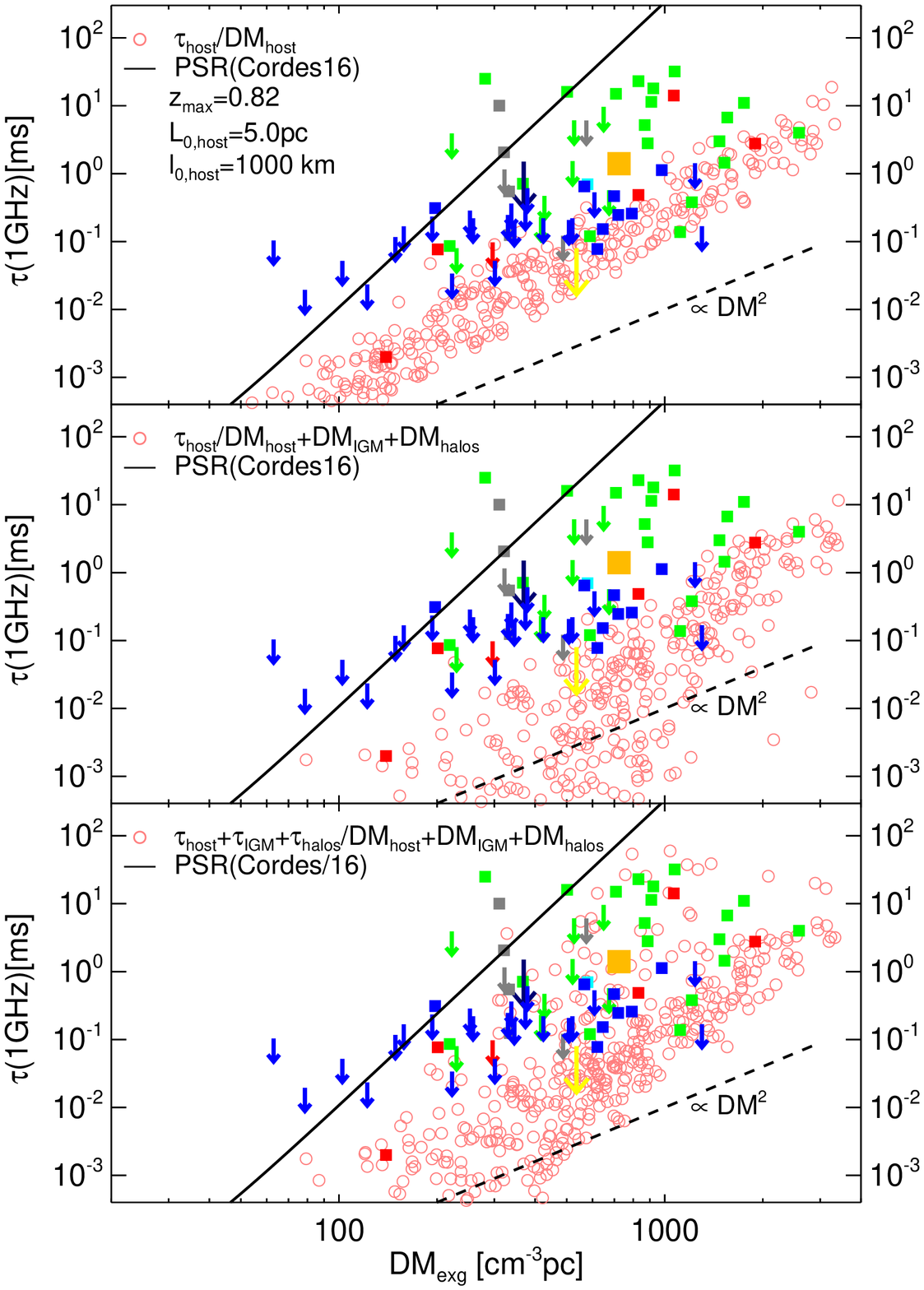}
}
\caption{Same as Fig.~\ref{fig:dm_tau_modela}, (a): results for Model-C ; (b): results for Model-D.}
\end{center}
\label{fig:dm_tau_modelbc}
\end{figure*}

\subsection{DM-$\tau$ relation, contributions to $\tau$}
The pink circles in Fig.~\ref{fig:dm_tau_modela}(a) show the extragalactic DM-$\tau$ relations of 500 randomly selected mock sources adopting Model-A, against to the observed events that either have reported $\tau$(solid symbols) or upper limiter of $\tau$(downward arrows). The outer scale in three types of medium are $L_0=5.0$pc. In Fig.~\ref{fig:dm_tau_modela}(a), each panel indicates the contribution to the extragalactic DM and $\tau$ of mock source by a particular component, or by the sum of two/three components. In the bottom-right panel, where both the DM and $\tau$ of mock sources are the sum of all the three types of intervening medium, we can see that the distribution of mock sources in the DM-$\tau$ space basically overlaps with the observed events. Most of the sources are below the DM-$\tau$ relation of pulsars in the MW to some extent(\citealt{2016arXiv160505890C}). Comparing the bottom-right panel to the other five panels in Fig.~\ref{fig:dm_tau_modela}(a), we conclude that every components of the intervening medium are important ingredients to explain the observed DM-$\tau$ relation.  

The results of Model-B, Model-C and Model-D are presented in Fig.~\ref{fig:dm_tau_modela}(b) and Fig.~\ref{fig:dm_tau_modelbc}. In both of Model-B and Model-C, the relative importance of the IGM to $\tau$ is enhanced, as the inner scale of turbulence in the IGM now equals to that of host and foreground halos. In comparison to Model-A, the improvement on $\tau_{\rm{IGM}}$ in Model-B can relax the requirement on the out scale $L_0$ up to $\sim 10$ pc in order to match the observation. However, in Model-C a much shorter $L_0$, i.e, $\sim 0.2$ pc, is needed to reproduce the observed DM-$\tau$ relation if the inner scale $l_0$ is $1$ AU in all the intervening medium. For Model-D, the inner and outer scale of turbulence in the host are set to the same values with Model-A. The foreground halos can still cause significant scattering time of mock events as large as tens of milliseconds, if $\rm{L_{0,halos}}=50\, \rm{pc}$ and $\rm{l_{0,halos}}=10^6\, \rm{km}$. The IGM, however, can only cause a scattering time smaller than $0.1$ milliseconds at 1 GHz for most mock source, if the inner and outer scale of turbulence in the intergalactic region are 1 AU and 500 pc respectively.

\begin{figure*}[htbp]
\begin{center}
\hspace{-0.5cm}
\includegraphics[width=0.45\textwidth, trim=0 10 0 10,clip]{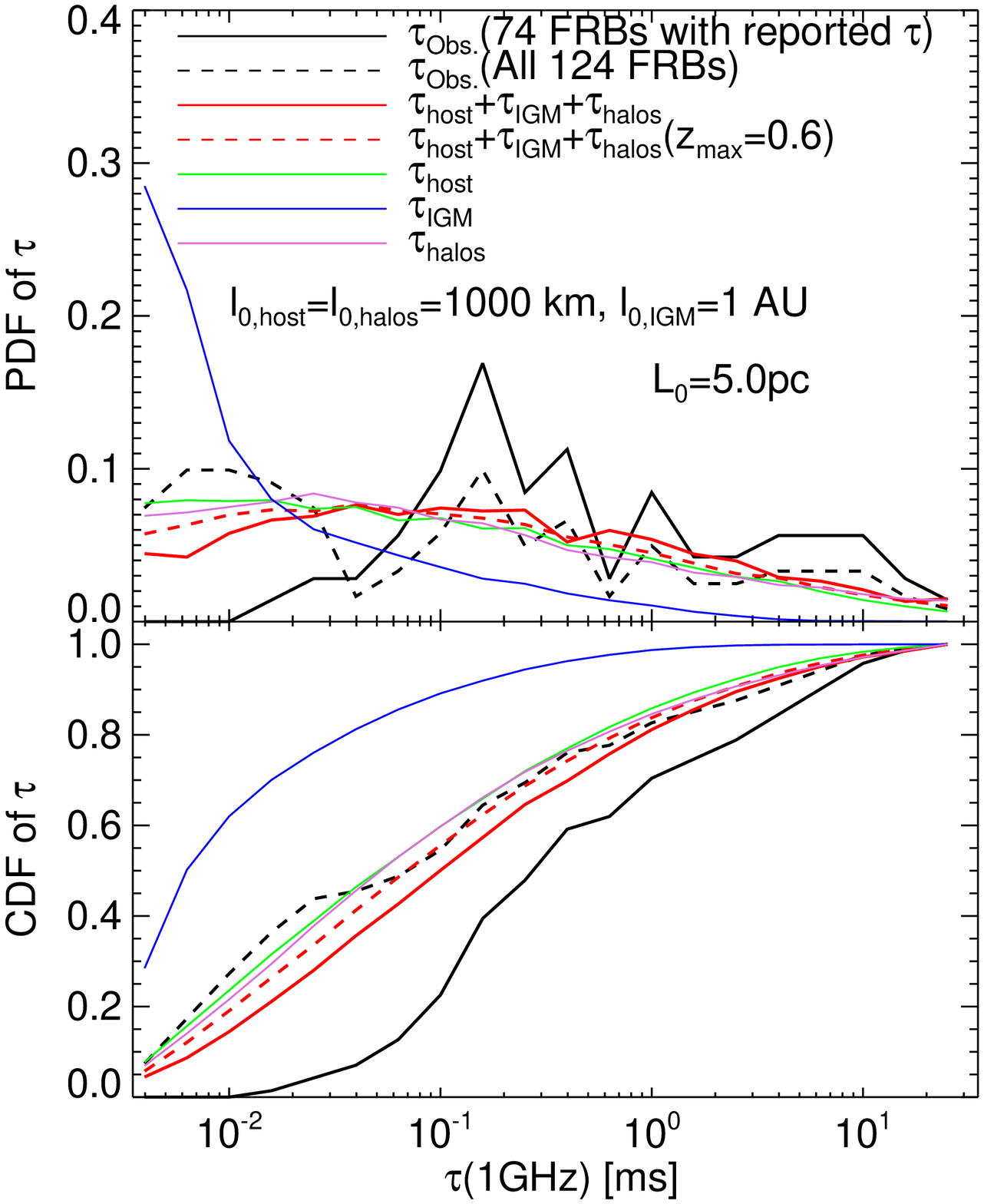}
\includegraphics[width=0.45\textwidth, trim=0 10 0 10,clip]{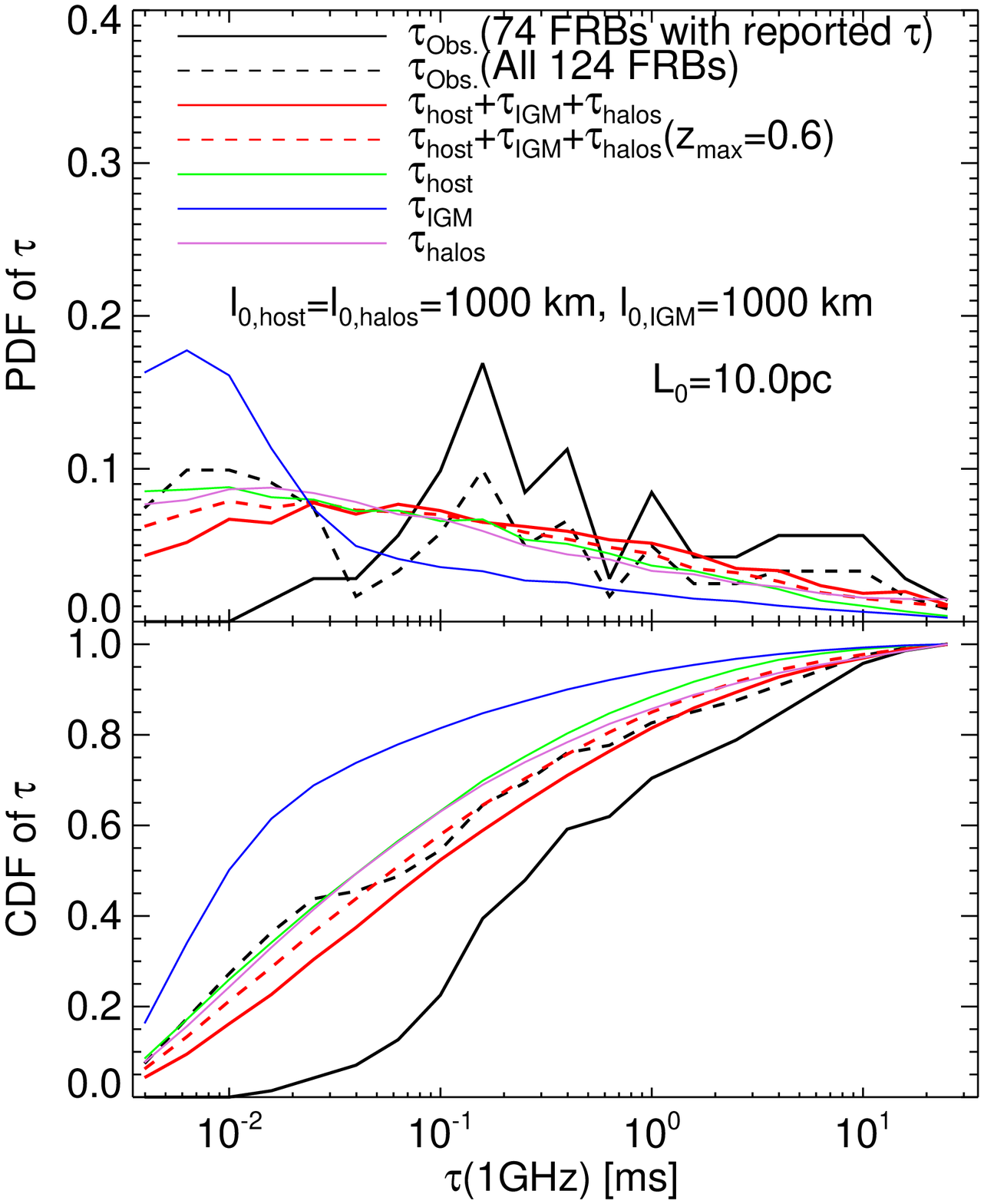}
\includegraphics[width=0.45\textwidth, trim=0 10 0 10,clip]{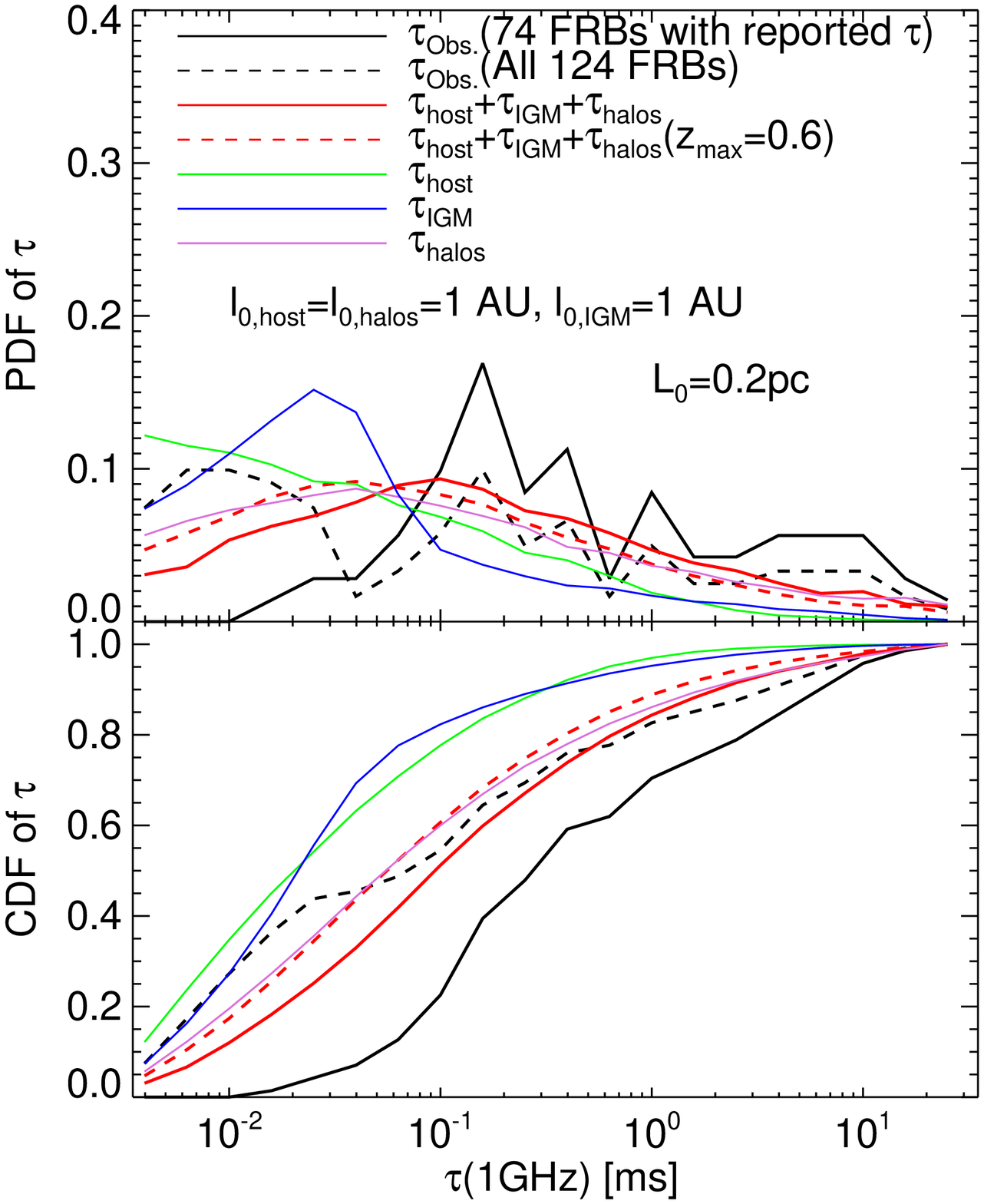}
\includegraphics[width=0.45\textwidth, trim=0 10 0 10,clip]{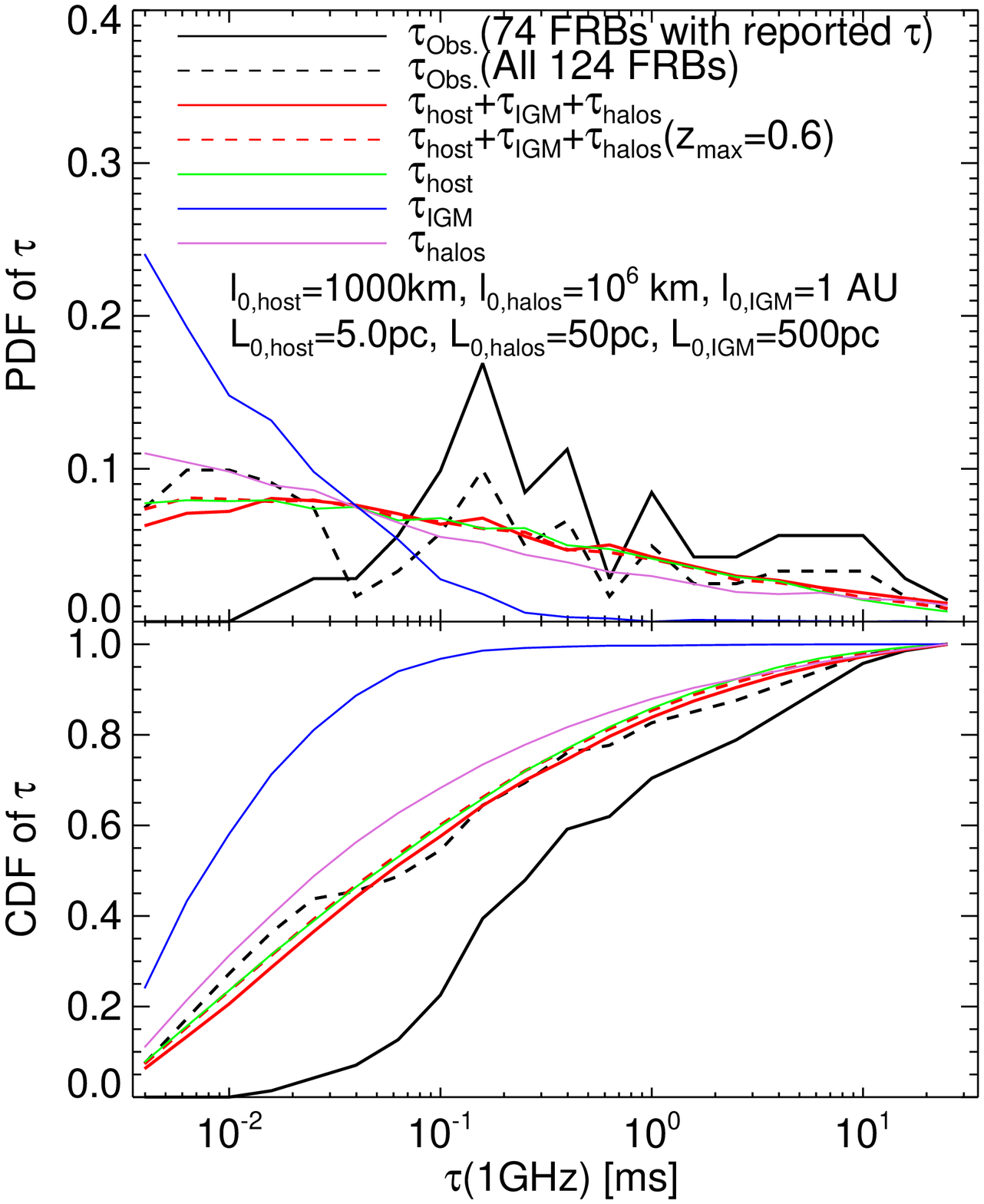}
\caption{Probability distribution function and cumulative distribution function of $\tau$ of mock samples, against observations(black solid and dashed lines, see the text for detail). Top-left, top-right, bottom-left and bottom-right panels show the results in Model-A, Model-B, Model-C and Model-D respectively.}
\end{center}
\label{fig:dm_tau_pdf}
\end{figure*}

\begin{figure*}[htbp]
\begin{center}
\hspace{-0.5cm}
\includegraphics[width=0.45\textwidth, trim=0 20 0 20,clip]{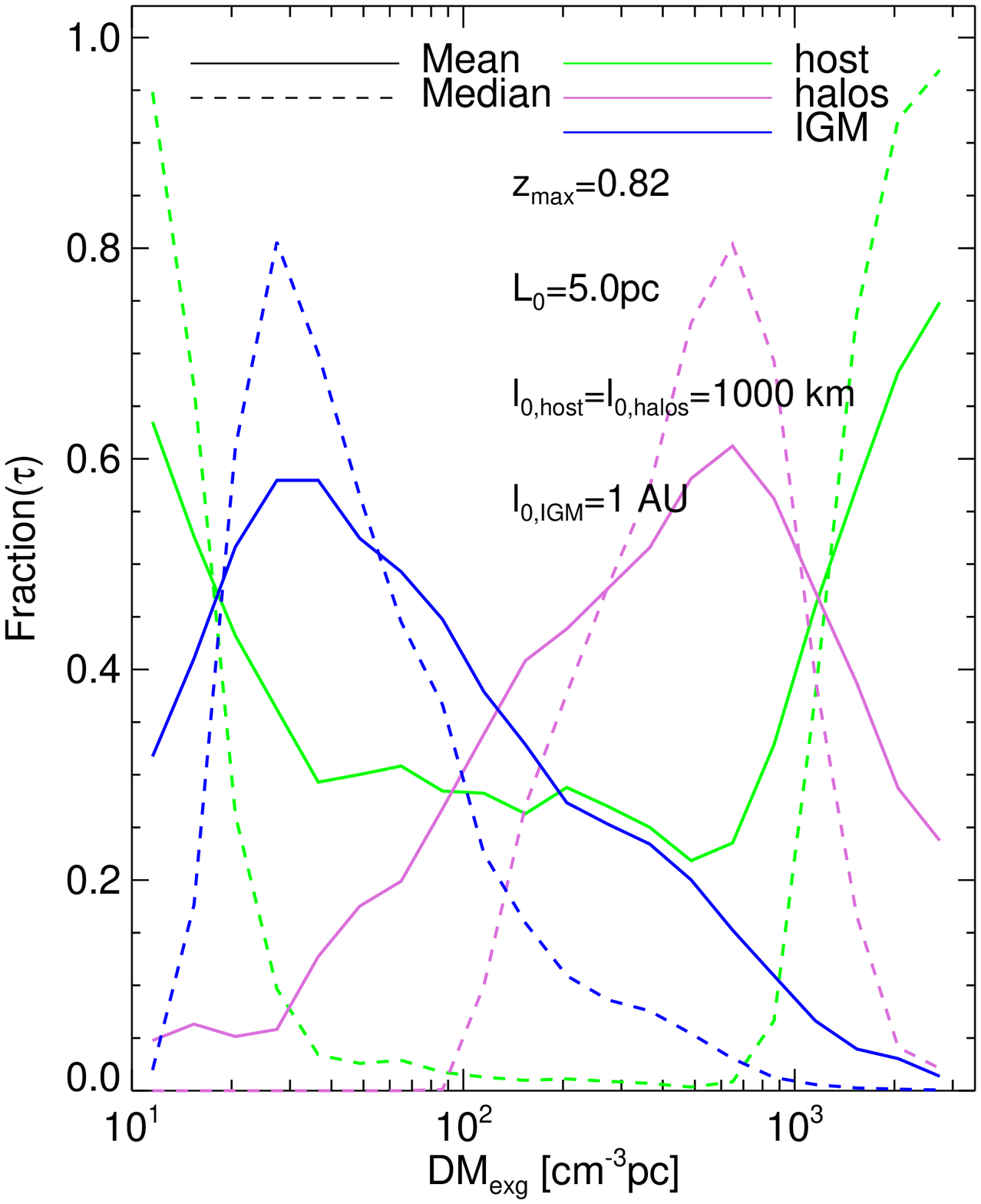}
\includegraphics[width=0.45\textwidth, trim=0 20 0 20,clip]{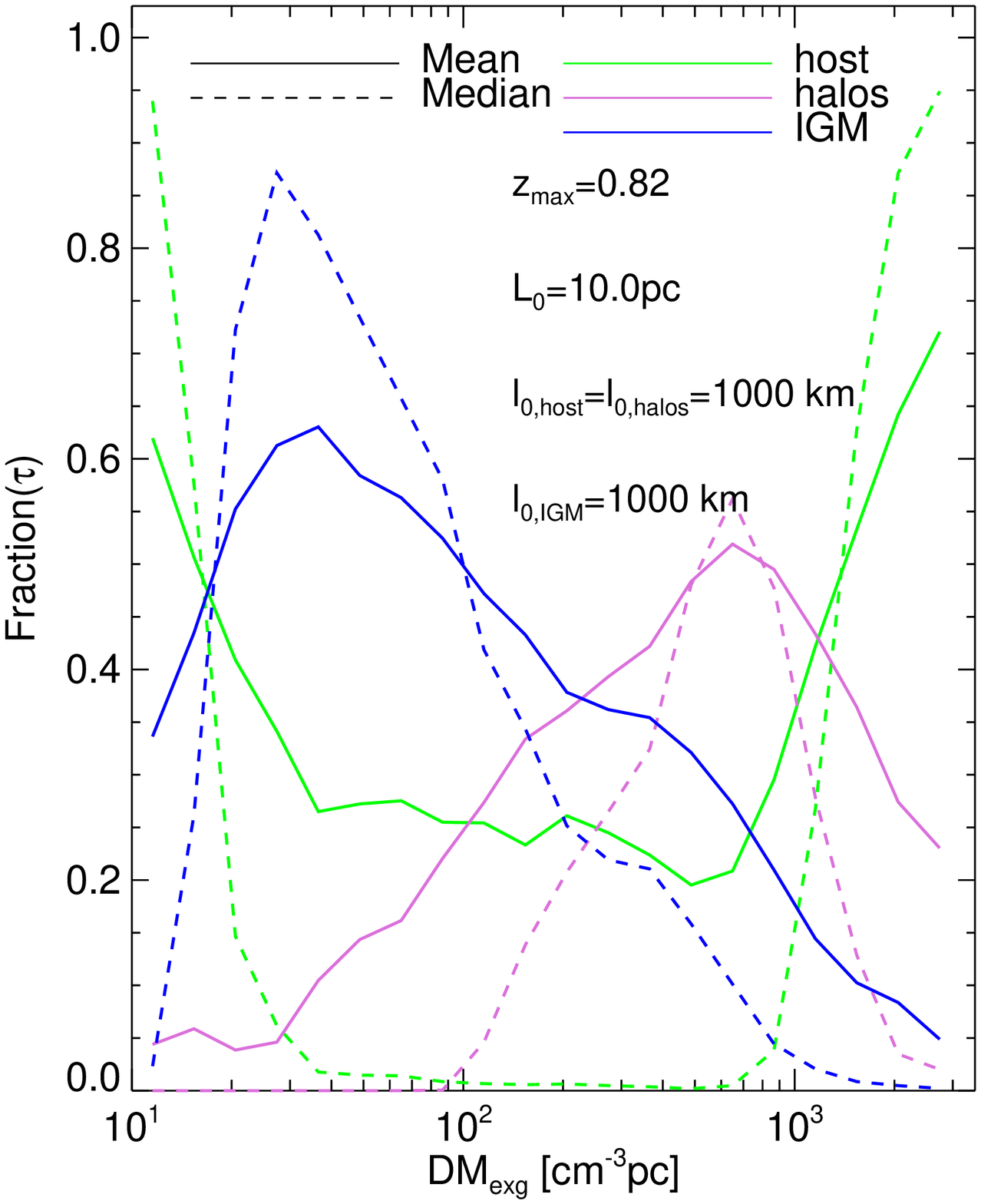}
\includegraphics[width=0.45\textwidth, trim=0 20 0 20,clip]{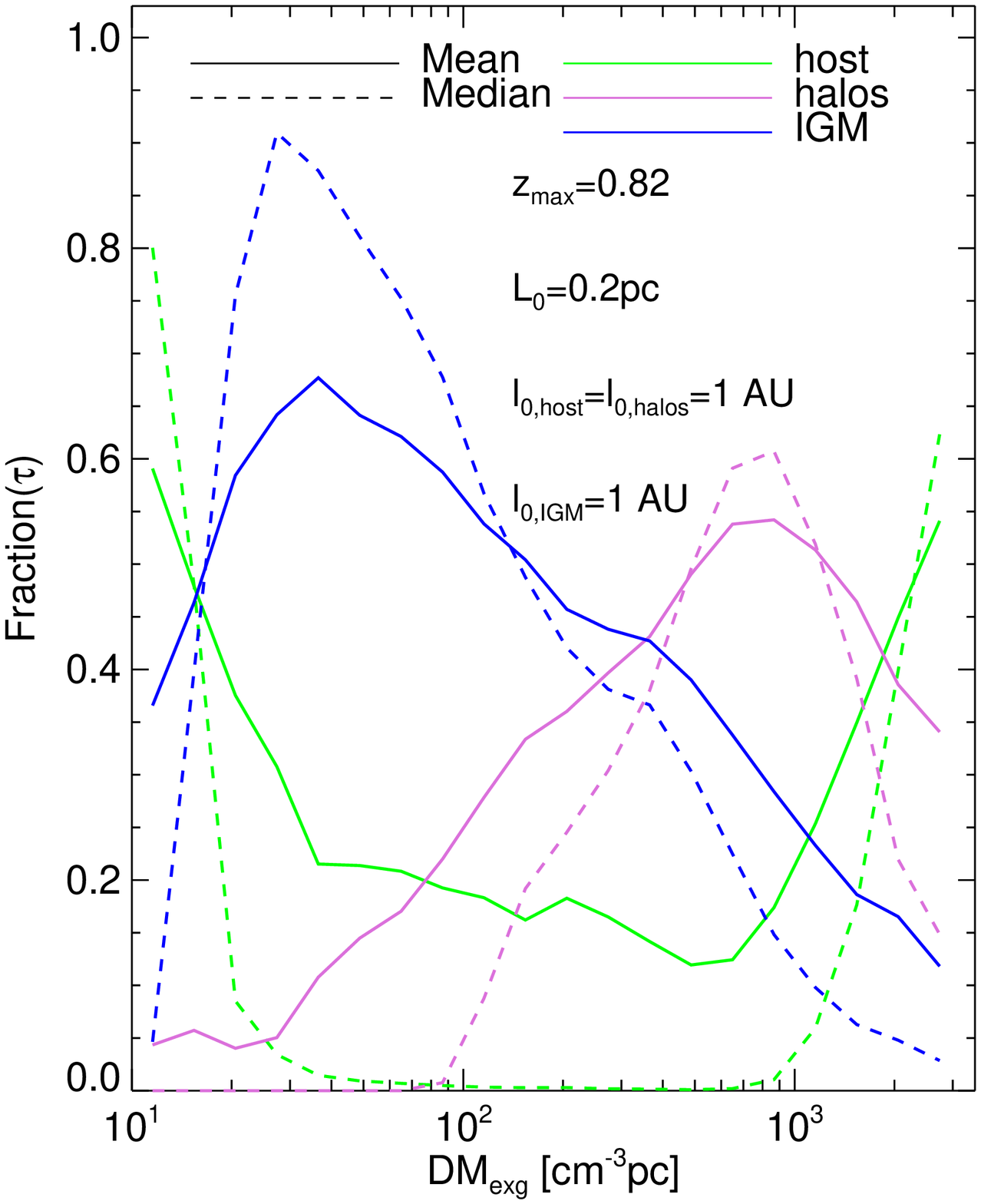}
\includegraphics[width=0.45\textwidth, trim=0 20 0 20,clip]{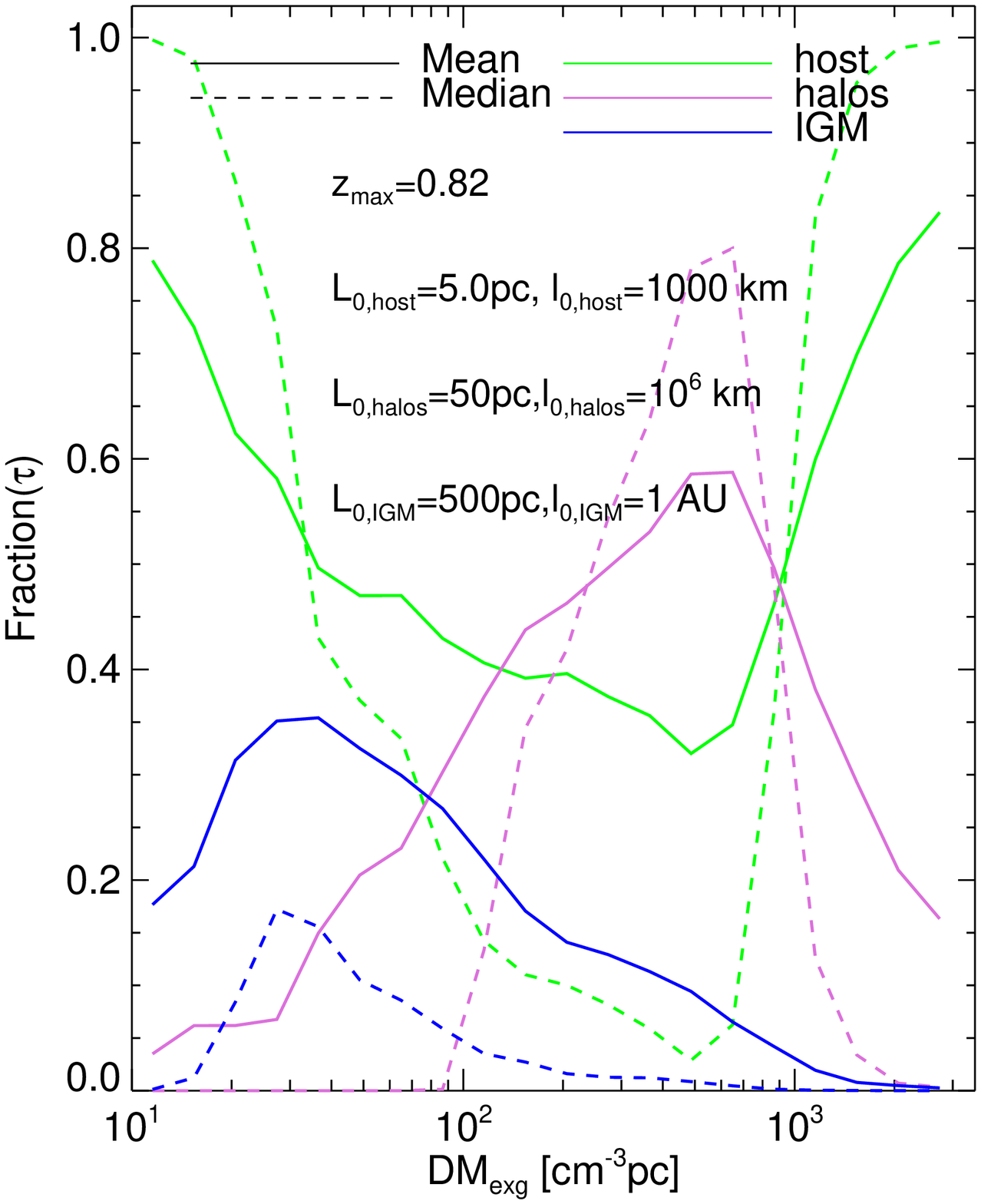}
\caption{The mean(solid) and median(dashed) fractions of $\tau$ of mock sources caused by the IGM(blue), foreground halos(purple) and the host halos(green) as a function of the total extragalactic DM. Top-left, top-right, bottom-left and bottom-right panels show the results in Model-A, Model-B, Model-C and Model-D respectively.} 
\end{center}
\label{fig:dm_tau_frac_model}
\end{figure*}

Fig.~\ref{fig:dm_tau_pdf} shows the probability density
function(PDF) and cumulative density functions of $\tau$ in four models, in comparison with the observed events. The $\tau$ distribution of mock sources is somewhat similar to that of the 74 observed events with reported $\tau$(or up limit) at $\tau>0.03\, \rm{ms}$. However, there are two major differences. First, the $\tau$ distribution of observed events show notable peaks and dips, which should partly because of fluctuations due to limited sample size. Second, there are many mock sources with very small $\tau$. Note that, there are 50 observed FRB events have neither a measured $\tau$ or a upper limit on $\tau$. Some of these events may have scattering of time scale a few to tens of $\mu$s, which needs observations with high time resolution to resolve(e.g. \citealt{2018MNRAS.478.1209F}; \citealt{2020MNRAS.497.3335D}). We find that if we hypothetically assign a small $\tau$, randomly distributed in the range $10^{-2.5}-10^{-1.5} \rm{ms}$, to each of these 50 events and include them in the $\tau$ distribution, the discrepancy between mock sources and observation on $\tau$ distribution would be largely relieved. 

On the other hand, Fig.~\ref{fig:dm_tau_pdf} indicates that in Model-A and Model-B, the contribution from the foreground halos to the scattering time is comparable to that of host halos, and are much stronger than that of the IGM. In Model-C, foreground halos dominate the time broadening of mock sources, while the contribution from host halos is comparable to the IGM. In Model-D, the foreground halos remains an important contributor to the scattering time. The IGM, however, plays a very limited role in the scattering.  In Z18, where the role of foreground halos is not separated from the diffuse IGM, we found the medium residing in cosmic knots and filaments may induce significant scattering. With improved simulation resolution and tools, our models in this work suggest that foreground halos may play important role in the scattering of FRBs. 

Finally, we measure the relative importance of the IGM, foreground halos and host halos to the scattering time $\tau$ in our models. Fig.~\ref{fig:dm_tau_frac_model} shows the mean and median fractions of $\tau$ contributed by the different medium as a function of the total extragalactic DM. Similar to the case of $\rm{DM_{exg}}$, the host halos dominates the $\tau$ for sources with $\rm{DM_{exg}}<20 {\rm{pc\, cm^{-3}}}$ or $\rm{DM_{exg}}>800-1000 {\rm{pc\, cm^{-3}}}$. The secondary contribution to $\tau$ is made by the IGM and foreground halos for sources with $\rm{DM_{exg}}<20 {\rm{pc\, cm^{-3}}}$, and $\rm{DM_{exg}}>800-1000 {\rm{pc\, cm^{-3}}}$ respectively.

For sources with total extragalactic DM in the range $20-800 {\rm{pc\, cm^{-3}}}$, the mean fraction of $\tau$ contributed by the IGM declines monotonously from $\sim 35-60\%$ at $\rm{DM_{exg}}=20 {\rm{pc\, cm^{-3}}}$ to $\sim 5-20\%$ at $\rm{DM_{exg}}=800 {\rm{pc\, cm^{-3}}}$ in the four different models considered here. In contrast, the contribution to $\tau$ from foreground halos increases monotonously from $5\%$ to $50-60\%$, while those from host halos decrease slowly from $40-55\%$ to $15-20\%$. For sources with with $20 \lesssim \rm{DM_{exg} \lesssim 200 pc\, cm^{-3}}$, the IGM is the primary contributor to $\tau$ in the first three models, but the host dominates in Model-D. While for sources with $200 \lesssim \rm{DM_{exg} \lesssim 800 pc\, cm^{-3}}$, the foreground halos are the primary contributors to the time scattering in all of the \textbf{four} models. However, the secondary contributor varies from model to model, depending on the values of $l_0$ and $L_0$.

Due to the same reason stated in section 4.1, our mock sources with $\rm{DM_{exg}} \geq 800 {\rm{pc\, cm^{-3}}}$ are biased by those with relatively large $\rm{DM}_{host}$ arising from a limited value of $z_{max}$. The, if $z_{max}$ was increased, the foreground halos would be expected to contribute more than $60\%$ of the scattering time at $\rm{DM}\geq 800 {\rm{pc\, cm^{-3}}}$, based on our models. Meanwhile, the average contribution from host halos to $\tau$ may drop below $\sim 15-20\%$, and keep declining slowly with the total DM increasing. Note that, these fractions only give the average values. There could be significant variations between different sources. Once again, we should remind the reader that the turbulent scales in the foreground halos and the IGM are barely known so far. Given recent progresses made in the study of CGM, Model-D might be more favorable currently. If the inner and outer scale of turbulence in the foreground halos and IGM are much larger than the assumed values in our models, the relative importance of foreground halos and IGM on scattering will be further decreased.

\section{Summary and Discussions}
Based on the high resolution cosmological AMR hydrodynamic simulation, we conduct an investigation on the dispersion and scattering measures induced by the host halos, the foreground halos and the IGM along the lines-of-sight to FRB events. We further produce a large number of mock samples of FRB events, considering the propagation effects of those intervening medium. We study the dispersion measure of these mock source. We investigate the scattering of mock source with different models on the turbulent scales in different medium. One of our goals is to justify whether the DM distribution and DM-$\tau$ relations inferred from the observed events can be reproduced by our mock samples. Particularly, we probe the DM-redshift relation, and offer a fitting formula that can be applied to yield a rough estimation of redshifts of FRB events from their dispersion measures $\rm{DM_{exg}}$.  In addition, we carry out a statistical analysis to estimate the relative importance of these three kinds of intervening medium giving rise to the DM and $\tau$ of FRB sources in our models. We summary our findings as follow:
\begin{enumerate}
\item  The median value of DM caused by foreground halos, $\rm{DM_{halos}(z)}$, is about 30\% of that caused by the IGM, $\rm{DM_{IGM}(z)}$. The median $\rm{DM_{halos}}$ are about 10, and 60 ${\rm{pc\, cm^{-3}}}$ at $z=0.1$ and $z=0.8$ respectively. However, the variance in $\rm{DM_{halos}(z)}$ is larger than that in $\rm{DM_{IGM}(z)}$ by a factor of 3-4, gradually increasing from about 100 ${\rm{pc\, cm^{-3}}}$ at $z=0.1$ to about 300 ${\rm{pc\, cm^{-3}}}$ at $z=0.8$. On the other hand, foreground halos contributes about $95\%$ of the scattering measure caused by the IGM and foreground halos. Improved simulation resolution would increase the variance in DM and SM caused by the IGM and foreground halos, mainly because of better resolved gas distribution in foreground halos. 

\item The inhomogeneous gaseous medium in host halos would lead to significant variations on the $\rm{DM_{host}}$ and $\rm{SM_{host}}$ along various radial trajectories in different host halos. $\rm{DM_{host}}$ spans a wide range of 1 to $3000 \ {\rm{pc\, cm^{-3}}}$, and deviates from a log-normal distribution, exhibiting an almost even distribution in the range $\sim 3-30 \  {\rm{pc\, cm^{-3}}}$, and then developing a bump peaked at $\sim 300 \  {\rm{pc\, cm^{-3}}}$.  This distribution shows minor evolution with time at $z\lesssim 1$, with the mean value increase slightly with increasing redshift. The median value of $\rm{DM_{host}}$ is about $100 \ {\rm{pc\, cm^{-3}}}$ at $z=0$.

\item Accounting the contributions to dispersion and scattering from the IGM, foreground halos and host halos simultaneously, we generate 50000 mock sources that are evenly distributed in the range $0.0<z<0.82$. The DM distribution of our mock sources agrees well with the observed events. Under the assumption that the inner scale of turbulence in the foreground halos, host halos and the IGM varies from 1000 km to 1AU and the outer scale varies from 0.2 to 10 pc, the distribution of our mock sources in the DM-$\tau$ space can broadly match up with the observations, and so does the probability distribution of the time scattering of mock sources also.

\item The fitting formula of median extragalactic dispersion - redshift (DM-z) relation extracted from our mock sample, i.e., Eqn.~\ref{eq:dm_all_z_fit},  can provide a rough estimation of redshifts of observed FRB events. Without any prior knowledge of the hosts, this relation enable us to recover the redshifts of 8 localized FRB events with errors $\delta z \lesssim 0.15$. Alternatively, by assuming $\rm{DM_{host}=100 pc\, cm^{-3}}$, the redshifts of those 8 events can also be recovered, with error $\delta z \lesssim 0.15$, from the median value of $\rm{DM_{halos}+DM_{IGM}}$ of our mock sources, i.e., Eqn.~\ref{eq:dm_z_fit}. For 5 of those 8 events, the errors of estimated redshift are around $\delta z \sim 0.05$. These errors are primarily result from the uncertainty of $\rm{DM_{host}}$. The uncertainty of dispersion induced by foreground halos are the secondary cause of errors. 

\item Statistically, the host halos dominates both the DM and $\tau$ for our mock sources at low dispersion ends, i.e., $\rm{DM_{exg}}\lesssim 30 {\rm{pc\, cm^{-3}}}$. Above which, the IGM makes a dominant contribution to DM. The mean fraction of $\rm{DM_{exg}}$ caused by IGM and host halos are $\sim 60\%$ and $\sim 35-45\%$ respectively at $\rm{DM_{exg}}=30 {\rm{pc\, cm^{-3}}}$, and declines to $\sim 55\%$ and $\sim 30\%$ at $\rm{DM_{exg}}\sim 800 {\rm{pc\, cm^{-3}}}$. While for the scattering time $\tau$, the relative importance varies in our different models. The host and foreground halos may be the most important scattering medium, while the IGM plays a limited role. The mean fraction of $\tau$ contributed by the IGM and host halos declines from $\sim 35-60\%$ and $\sim 40-55\%$ at $\rm{DM_{exg}}=30 {\rm{pc\, cm^{-3}}}$ to $\sim 5-20\%$ and $\sim 15-20\%$ at $\rm{DM_{exg}}=800 {\rm{pc\, cm^{-3}}}$ respectively. In the same range, the contribution to $\tau$ from foreground halos increases from $5\%$ to $50-60\%$. On average, the foreground halos could dominate the scattering time of FRB events with $\rm{DM_{exg}} \gtrsim 200-300 {\rm{pc\, cm^{-3}}}$ in our models. These trends are expected to hold even extending to larger $\rm{DM_{exg}}$.
\end{enumerate}

Following the threads in previous studies(e.g, \citealt{2013ApJ...776..125M}; \citealt{2014ApJ...780L..33M}; \citealt{2015MNRAS.451.4277D}; \citealt{2016arXiv160505890C}; \citealt{2018ApJ...865..147Z}), especially the models analyzed in \cite{2016arXiv160505890C}, our work can reproduce the observed DM distribution of FRBs in a self-consistent way, and explain their DM-$\tau$ relation with plausible models and assumptions on the turbulence state of intervening medium. Moreover, our work provides an overall quantitative analysis on the relative importance of the IGM, foreground halos and host halos on the dispersion measure and scattering time of FRB events based on our simulation and models. Our statistical results(e.g. Eqn. \ref{eq:dm_z_fit} and \ref{eq:dm_all_z_fit}) can be applied to approximately estimate the redshifts of FRBs sources with or without prior knowledge of their hosts, if the extragalactic DM is available. However, due to the significant variations, the derived redshift for a particular source should be treated with some cautions.   

Our study suggests that the IGM and host are the primary and secondary contributors respectively to the dispersion of FRB events with $\rm{DM_{exg}>30 pc\, cm^{-3}}$. The contribution from foreground halos to dispersion increases with $\rm{DM_{exg}}$ increasing.  The case is reversed for the scattering measure of FRB events. As demonstrated before, the foreground halos could also play important role in the scattering, and can be statistically the primary contributor to $\tau$ for sources with large extragalactic DM, if the inner and outer scales of turbulence in the foreground halos are close to, or about ten times of those in the host halos.This result agrees with the speculation in \cite{2013ApJ...776..125M}, where the intracluster medium is expected to dominate the scattering of high redshift FRB sources. Actually, the latest observations have suggested that the lines-of-sight toward a few FRBs should have passed through the gaseous halos of foreground galaxies(e.g. \citealt{2019Sci...366..231P}). Accordingly, the gaseous halos of foreground galaxies can indeed be a possible location where the FRB signals go through significant scattering. The FRB events provide a potential tool to probe the circumgalactic medium of foreground galaxies and the baryonic cosmic web, on account of very high events rate of FRBs predicted theoretically.

It is noticed that our results are based on some important assumptions. Firstly, the baryonic gas is assumed to be fully ionized. In the real universe, some gas would remain neutral and thus the ionization rate would small than 1.0. Nevertheless, this effect may have been largely canceled out by the relatively lower gas fraction in our simulation, due to over produced stellar mass. The second assumption is made for the turbulent state of these medium, in particular, regarding the density power spectrum and the values of the inner and outer scales. If the real density distribution in these regions deviate from the Kolmogorov description(e.g. see \citealt{2016ApJ...832..199X} for some other models), the results about the contribution to the scattering by the three types of medium should be revised. 

As demonstrated in section 4, the values of the inner and outer scale can affect the scattering time significantly in different models. The values we adopted in the host halos may be plausible (e.g., \citealt{2016arXiv160505890C}). However, so far there is not any solid constraint on the turbulent state of the CGM and IGM below tens of kpc. Our Model-D provides an example for cases that the turbulent scales in the foreground halos and IGM are larger than that in the host halos. In this case, the foreground halos may still be an important contributor to the scattering, but the IGM plays a very limited role in the scattering. 


Addition attention needs to be drawn to the redshift cutoff of our mock source at $z=0.82$, and the assumption of the uniform redshift distribution . We put an redshift cut at $0.82$ by hand, because the limited number of snapshots generated by our cosmological simulation. In reality, it is very likely that there are FRB events at redshift higher than 0.82. Including more events at redshifts higher than $0.82$ would alter more or less the statistical results of mock sources with large $\rm{DM_{exg}}$. Nevertheless, we expect the trends regarding the relative importance of different medium on the dispersion and scattering of FRBs in the range $\rm{DM_{exg}}=30-800 {\rm{pc\, cm^{-3}}} $ can be extended to the regime with $\rm{DM_{exg}}>800 {\rm{pc\, cm^{-3}}} $. On the other hand, we have assumed an uniform redshift distribution for the sake of simplication. If the redshift distribution of FRBs deviates from a even distribution,which is also very likely, our results regarding the probability distribution function of DM and $\tau$ may be also revised. 

In addition, the role of local medium surrounding the FRB source is not considered in our models. Also, our results may have underestimate the roles of host halos and foreground halos, due to a limited spatial resolution in our simulation. Moreover, different sub-grid models of stellar feedback and AGN feedback can lead to different gas distribution in the host and foreground halos(e.g \citealt{2019BAAS...51c.297B}). Further investigations in the future are needed to estimate the effects of these factors. Meanwhile, more observed events with high resolution spectra, and knowledge of their host galaxy and foreground halos coming in the future will help to determine the relative contribution of DM and scattering from the progenitor environment, the host ISM, foreground halos and the diffuse IGM(e.g. \citealt{2019Sci...366..231P}; \citealt{2020MNRAS.497.3335D} ).

\section*{Acknowledgements}
We thank the anonymous referees for their useful comments on our manuscript. 
This work is supported by the Key Program of the National Natural Science Foundation of China (NFSC) through grant 11733010. W.S.Z. is supported by
the NSFC grant 11673077. F.L.L. is supported by the NSFC grant 11851301. The cosmological hydrodynamic simulation was run on the Tianhe-II supercomputer. The post-simulation analysis is carried on the HPC facility of School of Physics and Astronomy, SYSU.  





\bibliography{ref}




\end{document}